\documentclass{aa}  
\usepackage{graphicx}
\usepackage{txfonts}
\usepackage{natbib}
\bibpunct{(}{)}{;}{a}{}{,} % to follow the A&A style

\def\aaps{A\&AS}
\def\aap{A\&A}
\def\apj{ApJ}
\def\apjs{ApJS}
\def\aj{AJ}

\def \hi {\ion{H}{i}}

\def\kms{km\,s$^{-1}$}

\def\deg{\hbox{$^\circ$}}
\def\arcmin{\hbox{$^\prime$}}

\def\fdg{\hbox{$.\!\!^\circ$}}
\def\farcm{\hbox{$.\mkern-4mu^\prime$}}

\defcitealias{Naomi2009}{Paper~I}

\begin{document}

   \title{GASS: The Parkes Galactic All-Sky Survey. }

   \subtitle{II. Stray-Radiation Correction and Second Data Release}

\author{P.\ M.\ W.\ Kalberla,\inst{1} 
N.\ M.\ McClure-Griffiths,\inst{2} 
D.\ J.\ Pisano,\inst{3,10} 
M.\ R.\ Calabretta,\inst{2} 
H.\ Alyson Ford,\inst{2,5,9}
Felix J.\ Lockman,\inst{4}
L.\ Staveley-Smith,\inst{6}
J.\ Kerp,\inst{1}
B.\ Winkel,\inst{1}
T.\ Murphy,\inst{7,8}
and 
K.\ Newton-McGee\inst{2,8}   } 

\institute{Argelander-Institut f\"ur Astronomie, Universit\"at Bonn, Auf
  dem H\"ugel 71, 53121 Bonn, Germany,  
  \email{pkalberla@astro.uni-bonn.de, jkerp@astro.uni-bonn.de, 
   bwinkel@astro.uni-bonn.de} 
\and 
Australia Telescope National Facility, CSIRO, Marsfield
  NSW 2122, Australia 
  \email{naomi.mcclure-griffiths@csiro.au,mark.calabretta@csiro.au} 
\and 
Department of Physics, West Virginia University, P.O. Box 6315, 
Morgantown, WV 26506 
\email{djpisano@mail.wvu.edu}
\and
National Radio Astronomy Observatory, Green
  Bank, WV 24944 \email{jlockman@nrao.edu} 
\and 
Centre for Astrophysics and Supercomputing, Swinburne
  University of Technology, Hawthorn VIC 3122, Australia
  \and
International Centre for Radio Astronomy Research, M468, 
University of Western Australia, Crawley, WA 6009, Australia 
\email{Lister.Staveley-Smith@icrar.org}
\and
School of Physics, The University of Sydney, NSW 2006,
  Australia \email{tara@physics.usyd.edu.au} 
\and
School of Information Technologies, The University of
  Sydney, NSW 2006, Australia \email{katherine.newton-mcgee@defence.gov.au}
\and
Department of Astronomy, University of Michigan, Ann Arbor, MI 48109, USA 
\email{haford@umich.edu}
\and
Adjunct Assistant Astronomer at the National Radio Astronomy Observatory
}

   \authorrunning{P.\,M.\,W. Kalberla et al. } 

   \titlerunning{GASS II, Corrected for Stray-Radiation and Second Data
     Release}

   \offprints{P.\,M.\,W. Kalberla}

   \date{Received 28 December 2009 / Accepted 1 July 2010 }

% \abstract{}{}{}{}{} 
% 5 {} token are mandatory
  \abstract 
  {The Parkes Galactic All-Sky Survey (GASS) is a survey of Galactic atomic
    hydrogen (\hi) emission in the southern sky observed with the Parkes 64-m
    Radio Telescope. The first data release was published by McClure-Griffiths
    et al. (2009).}
  {We remove instrumental effects that affect the GASS and present the second
    data release.  }
  {We calculate the stray-radiation by convolving the all-sky response
    of the Parkes antenna with the brightness temperature distribution
    from the Leiden/Argentine/Bonn (LAB) all sky 21-cm line survey,
    with major contributions from the 30-m dish of the Instituto
    Argentino de Radioastronom\'ia (IAR) in the southern
    sky. Remaining instrumental baselines are corrected using the LAB
    data for a first guess of emission-free baseline regions. Radio
    frequency interference is removed by median filtering. }
  {After applying these corrections to the GASS we find an excellent agreement
    with the Leiden/Argentine/Bonn (LAB) survey. The GASS is the highest 
    spatial resolution, most sensitive, and is currently the
    most accurate \hi~ survey of the Galactic \hi~ emission in the southern sky.
  We provide a web interface for generation and download of FITS cubes.}
  {}
  % -------------------------------
% -------------------------------
  \maketitle
%
%________________________________________________________________

\begin{table*}
\begin{center}
\begin{tabular}{l|| c c c c c}
  \hline
  Parameter & LAB & HIPASS-HVC & GASS &  GALFA-HI & EBHIS\\
  \hline
  \hline
  Sky coverage & full sky & $\delta \la 2\deg$  & $\delta \la 1\deg$ & 
$-1\deg \la \delta \la 38\deg$ & $\delta \ga -5\deg$\\
  Average beam FWHM & 36\arcmin & 14.4\arcmin & 14.4\arcmin & 3.4\arcmin  & 9.5\arcmin \\
  LSR velocity range & $-450 < v < +400$ & $-700 < v < +1000$ &
  $-468< v < +468$ &  $-750 < v < +750$  &  $-1000 < v < +1000$\\
%  Channel separation & 1.03 \kms & 13.2 \kms & 0.82 \kms & 1.25 \kms \\
  Channel width & 1.27 \kms & 26.4 \kms & 1.0 \kms & 0.18 \kms & 1.25 \kms\\
  $1 \sigma T_B$ noise & 90 mK &   24 mK  & 57 mK & $ \la 100$ mK & $ \la 90$ mK\\
%  $3 \sigma N_{\rm HI}$ Sensitivity limit for $\Delta v = 30 $ \kms & $3.2~ 10^{16} {\rm cm}^{-2}$ &  $ 0.8~ 10^{16} {\rm cm}^{-2}$ &  $1.8~ 10^{16} {\rm cm}^{-2}$ \\
\end{tabular}
\caption[]{Survey parameters for major single dish surveys of the Galactic
  21-cm line emission. GALFA-HI and EBHIS are still in progress. }
\label{table1}
\end{center}
\end{table*}

\section{Introduction}

Atomic hydrogen (\hi) is ubiquitous, it traces the interstellar medium (ISM)
in disk galaxies over a broad range of physical conditions. Its 21-cm emission
line is a key probe of the structure and dynamics of galaxies over a broad
range of different densities and temperatures. \hi~ is the key element to
understand the evolution of the ISM in general \citep{ARAA2009}. 21-cm line
observations of the Milky Way are of particular interest because small scale
structures are observable which can not be resolved with observations of any
other galaxy. A number of notable items have been detailed by
\citet[][hereafter Paper I]{Naomi2009}, this includes the impact of massive
stars on the ISM, the disk-halo interaction, the ISM life-cycle and the
formation of cold clouds. \hi~ is a perfect tracer of the Galactic structure
and there is a good correlation between gas and dust \citep{Schlegel1998}.

Galactic \hi~ data are important for an exploration of the early universe
because of the photoelectric absorption caused by the ISM in the soft X-ray
energy range below 1\,keV. The attenuation of the X-rays by the Galactic
interstellar medium increases with decreasing X-ray energy
\citep{Kerp2003}. The emission of active Galactic nuclei at high redshifts ($z
\sim 10$) or the faint emission of clusters of galaxies at moderate redshifts
($3 \la z \la 5$) is shifted to this soft X-ray band. It is not possible to
analyze the X-ray data quantitatively without knowing in detail the
distribution of the Galactic interstellar medium across the target.

High resolution single dish data are needed but neither the \hi~ Parkes All
Sky Survey (HIPASS) by \citet{Barnes2001} nor the reprocessed version
(HIPASS-HVC) by \citet{Putman2002} provide accurate column densities for the
Galactic emission in the southern sky.  The GASS provides such data. For the
northern sky the Galactic Arecibo L-Band Feed \hi~ (GALFA-HI) survey
\citep{Peek2008} and the Effelsberg-Bonn \hi~ Survey (EBHIS) are still in
progress \citep{Winkel2010}. Major parameters for these surveys are compared
in Table \ref{table1}. Currently the LAB and the GASS are the only surveys
that are easily accessible for data mining.

The Galactic All-Sky Survey (GASS) is a 21-cm line survey covering the
southern sky for all declinations $\delta \la 1\deg$.  The observations were
made with the multibeam system on the 64-m Parkes Radio Telescope. The
intrinsic angular resolution of the data is $14\farcm4$ (FWHM). The velocity
resolution is 1.0 \kms~ and the useful bandpass covers a velocity range $ {\rm
  |v_{lsr}|} \la 468$ \kms~ for all of the observations; some data cover up to
${\rm |v_{lsr}|} \la 500$ \kms. The GASS is the most sensitive, highest
angular resolution large-scale survey of Galactic \hi~ emission ever made in
the southern sky. The first data release is available at
http://www.atnf.csiro.au/research/GASS.

A detailed description of the GASS project, focusing on the survey goals and
techniques has been given by \citetalias{Naomi2009}. In Paper I all of the
initial data reduction and imaging is described in detail. Here we focus on
the post-processing, in particular on corrections for stray-radiation,
instrumental baselines, and radio frequency interference (RFI). We will note
where our data reduction differs from that described in
\citetalias{Naomi2009}.

For 21-cm line observations there are a few unavoidable instrumental issues
that can potentially degrade the data. The first is that a radio telescope
does not have a perfect beam, so some radiation enters through the sidelobes
of the antenna, causing spurious emission features. The second effect is that
the receiver gain may drift, and it can suffer from instrumental spectral
structure (the instrumental baseline) that may be time-dependent.  These
influences need to be minimized, but the raw observations do not allow a clear
identification and separation of individual instrumental problems. For example,
spurious profile wings may be caused by stray-radiation but also by
instrumental baseline defects. Temporal fluctuations in the line emission may
be due to gain variations but may also be caused by spurious emission from the
sidelobes.

A strategy is needed to disentangle these instrumental issues.  For the
stray-radiation correction a basic prerequisite is the knowledge of the \hi~
brightness temperature distribution provided by the Leiden/Argentine/Bonn
(LAB) 21-cm line survey \citep{Kalberla2005}; most important in the context of
the GASS survey is the contribution from the 30-m dish of the Instituto
Argentino de Radioastronom\'ia (IAR) in the southern sky
\citep{Bajaja2005}. These surveys provide clean data with an angular
resolution of $\sim 36\arcmin$ and allow us to solve the Fredholm integral
equation which describes the deconvolution of the raw observations (antenna
temperatures) with respect to the influences of the antenna
response. Bootstrapping from the LAB survey we use a model of the antenna
response for the 13-beam receiver based on the known feed patterns and the
telescope geometry. Section \ref{SR} describes how spurious features from the
sidelobes are removed.

Next we remove baseline defects. For this the LAB survey is used for initial
estimates of velocity ranges that do not contain line emission. In
Sect. \ref{Baseline} we describe how baselines are refined in an iterative
way. A gain calibration requires some preliminary baseline corrections but
such a calibration also affects the accuracy of the corrections for
stray-radiation. The calibration is described in Sect. \ref{Cal}. All
corrections discussed in Sects. \ref{SR} to \ref{Cal} need to be iterated
several times.

RFI is increasingly a problem for sensitive observations, and post-processing
mitigation strategies are essential. We use flags which were set in the first
stage of the data processing by {\it Livedata} described in
\citetalias{Naomi2009} and iteratively remove the remaining interference as
discussed in Sect. \ref{RFI}. Throughout the reduction we take care that the
corrections described in Sects. \ref{SR} to \ref{Cal} are not affected by RFI.

In comparison with the LAB survey, the GASS represents an improvement in
velocity resolution and sensitivity, but most important is the enhanced
spatial resolution (Table \ref{table1}).  The generation of FITS images with
user-definable parameters is described in Sect. \ref{Images}. We discuss some
examples of the newly corrected data in Sect. \ref{Results} and complete the
paper with a description of a web interface for download (Sect. \ref{Data}).

\section{The stray-radiation correction}
\label{SR}

A major problem when observing 21-cm emission lines is caused by the fact that
Galactic emission is seen in all directions. Most prominent is the bright
Galactic plane which extends across the whole sky.  It is unavoidable that
stray 21-cm emission from the plane will get picked up by sidelobes of the
antenna.  The main sidelobes are the result of reflections from the feed legs
that carry the prime focus cabin.  Also radiation originating from regions
outside the rim of the reflector, the spillover region, is
received. Stray-radiation varies with time \citep{vanWoerden1962} and for
positions with very low \hi~the spurious emission features may amount to 50\%
of the total received signal \citep{Kalberla1980,Lockman1986,Hartmann1996}. In
some GASS spectra stray-radiation was found to comprise 35\% of the observed
emission.

Stray-radiation may not be neglected if we are interested in low-intensity
profile wings that may originate from lines with large velocity dispersion, or
in deriving accurate \hi~ column densities from 21-cm observations.  Observed
lines may originate either from the main beam, the ``real'' signal, or may
have been picked up from other directions, the ``stray'' signal.  Without
prior knowledge of instrumental issues it is also very difficult to
distinguish between faint line emission and instrumental baseline defects.

One way to minimize stray-radiation is by minimizing scattering surfaces in
the telescope aperture. The optimal case appears to be a telescope constructed
with a clean unblocked aperture \citep{Dickey1990}. Prime examples are the
Bell Labs horn reflector antenna \citep{Stark1992} and the Green Bank
Telescope (GBT; \citealt{Prestage2009}). A typical parabolic reflector antenna
has a main beam efficiency of 70\%, for unblocked apertures this improves to
$\ga 90$\%. Sidelobe contributions are thus reduced from 30\% to 10\% but are
still recognizable \citep{Lockman2005}. Depending on the observed position,
emission from the spillover region may cause severe contaminations in the case
of an unblocked aperture with a receiver in the secondary focus
\citep{Robishaw2009}.

A numerical solution of the stray-radiation problem was proposed by
\citet{vanWoerden1962}. He demonstrated convincingly that the stray component
can be determined by convolving the antenna pattern with the brightness
temperature distribution on the sky. Yet, neither the knowledge about sidelobe
structures, nor the computing power was sufficient enough for accurate
calculations in 1962. The first reliable calculations became available for the
Effelsberg telescope \citep{Kalberla1980,Kalberla1980s}, and later for the
resurfaced Dwingeloo dish \citep{Hartmann1996}, the 26-m Telescope at the
Dominion Radio Astrophysical Observatory \citep{Higgs2000}, and the 30-m
telescope at Villa Elisa \citep{Bajaja2005}. A refined correction of the
Leiden/Dwingeloo survey \citep{Atlas1997} in combination with the Villa Elisa
observations led to the LAB survey \citep{Kalberla2005}. Currently the LAB is
considered to be the most accurate Galactic all-sky survey.

\subsection{Basics }

We use a correction algorithm developed by \citet{Kalberla1980}. Antenna
temperatures $T_a$ observed by radio telescopes can be approximated in
Cartesian coordinates as a convolution of the true temperature distribution
$T$ on the sky with the beam pattern $P$ of the antenna
 
\begin{equation}
T_a(x,y)  =  \int P(x-x',y-y') T(x',y') dx' dy'.
\end{equation}
This is an approximation, but is sufficient to demonstrate the basics. In
general, $T_a$ is time and frequency dependent, spherical coordinates should
be used and the integration needs to be extended over the observable part of
the sky as well as the ground if there are reflections.  We do not include in
these equations the effects of atmospheric attenuation, though this must be
considered for emission entering both the main beam and the sidelobes
\citep{Lockman2002}. 

For the pattern $P$ of the antenna we use the normalization
\begin{equation}
\int P(x,y) dx dy  = 1
\end{equation}
in addition, we split the pattern into the main beam area (MB) and the 
stray-pattern (SP)

\begin{eqnarray}
  T_a(x,y) = \int_{MB} P(x-x',y-y') T(x',y') dx' dy' + \nonumber \\
  \int_{SP}P(x-x',y-y') T(x',y') dx' dy'.
\end{eqnarray}

Defining the main beam efficiency $\eta_{MB}$ of the telescope as 
\begin{equation}
\eta_{MB} = \int_{MB} P(x,y) dx dy 
\end{equation}
we may rewrite Eq. 3 as
\begin{equation}
T_B(x,y) =  \frac{T_a(x,y)}{\eta_{MB}}
- \frac{1}{\eta_{MB}}\int_{SP} P(x-x',y-y') T(x',y') dx' dy'.
\label{EQdeconvolve}
\end{equation}

Replacing $T_B$ by $T$ leads to a Fredholm integral equation of the second
kind which can be solved for $\eta_{MB} > 0.5$ \citep{Kalberla1980}.  $T_B$ is
the so-called brightness temperature, resulting from a convolution between the
main beam of the telescope and the true sky temperature $T$. A solution of
Eq. \ref{EQdeconvolve} is simplified considerably for a known all-sky
brightness temperature distribution $T$.  We use profiles from the LAB survey
as the currently best possible approximation to $T$. For a solution of
Eq. \ref{EQdeconvolve}, $P$ needs to be determined.  This is described in
Sects. \ref{nearSL} and \ref{farSL}.

\subsection{The correction algorithm }

The numerical correction algorithm is based on \citet{Kalberla1980}. It was
originally developed for the Effelsberg 100-m telescope, with some
improvements it was then applied to the LAB survey \citep{Kalberla2005}. We
modified this procedure for use with multi-beam systems, allowing individual
corrections for each of the beams. Due to the rotation of the offset receivers
with respect to the telescope structure, we take into account the variation of
each receiver's beam pattern on the sky with time.  To improve the accuracy of
the calculations we also changed the convolution algorithm for the inner part
of the antenna response pattern (within 6\deg~ of the main beam).  From the
LAB survey, a new and more accurate input sky for the convolution
(Eq. \ref{EQdeconvolve}) was generated.

\subsection{The multi-beam antenna pattern - near sidelobes }
\label{nearSL}

To solve Eq. \ref{EQdeconvolve} it is necessary to know the complete antenna
response $P$ for each of the 13 beams of the Parkes multi-feed system. In
practice there are severe limitations in measuring the sidelobes to the
required accuracy. We have therefore chosen to model the sidelobes. In
general, the far-field pattern of an antenna is the autocorrelation of the
aperture plane distribution. Using the measured feed response pattern and the
telescope geometry, including shadowing caused by the focus cabin and the feed
support legs, we derive the complex aperture distribution function for each
feed. The pattern can then be calculated by Fourier transformation.

\begin{figure}[!ht]
   \centering
   \includegraphics[angle=-90,width=5cm]{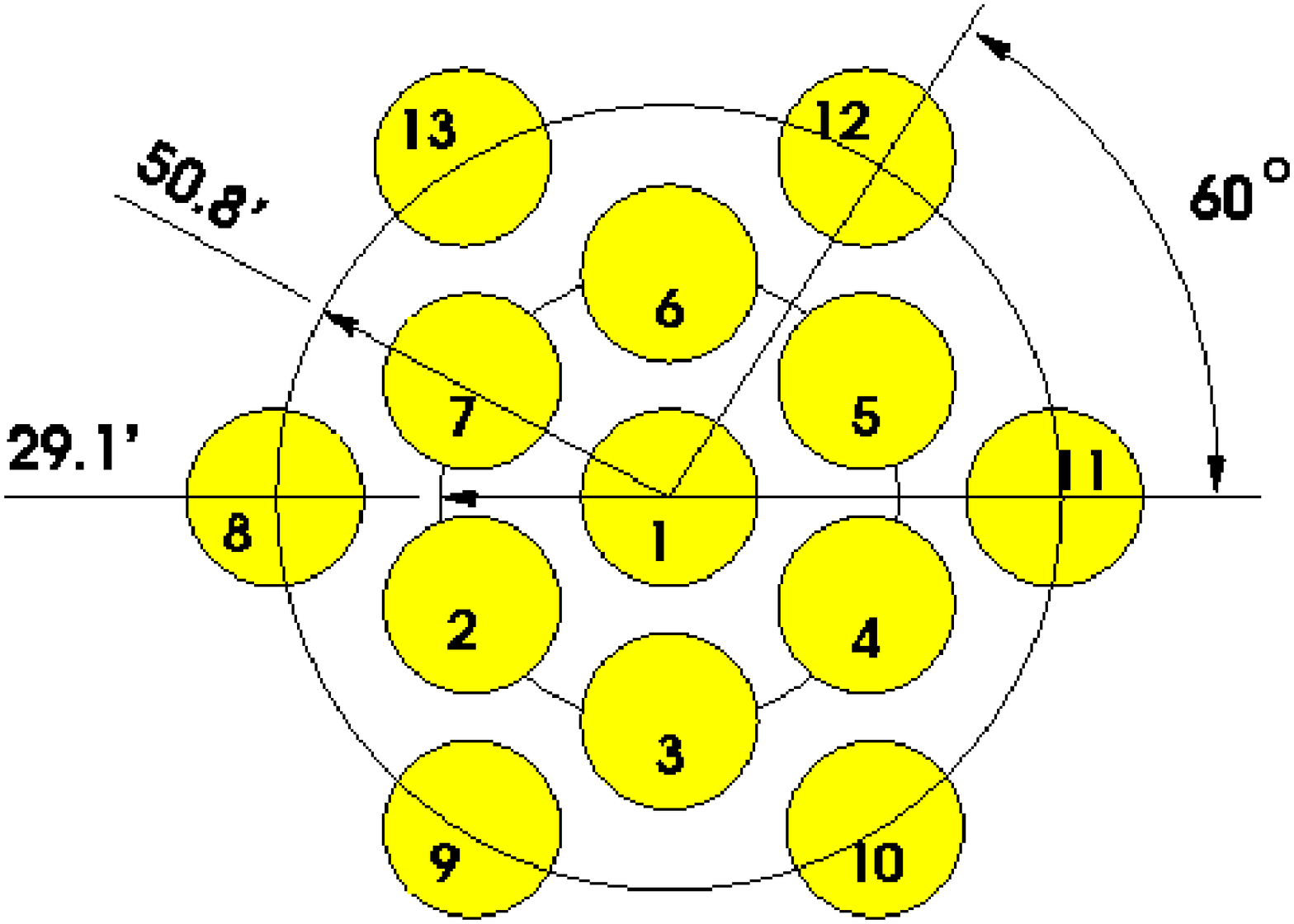}
   \includegraphics[angle=-90,width=6.cm]{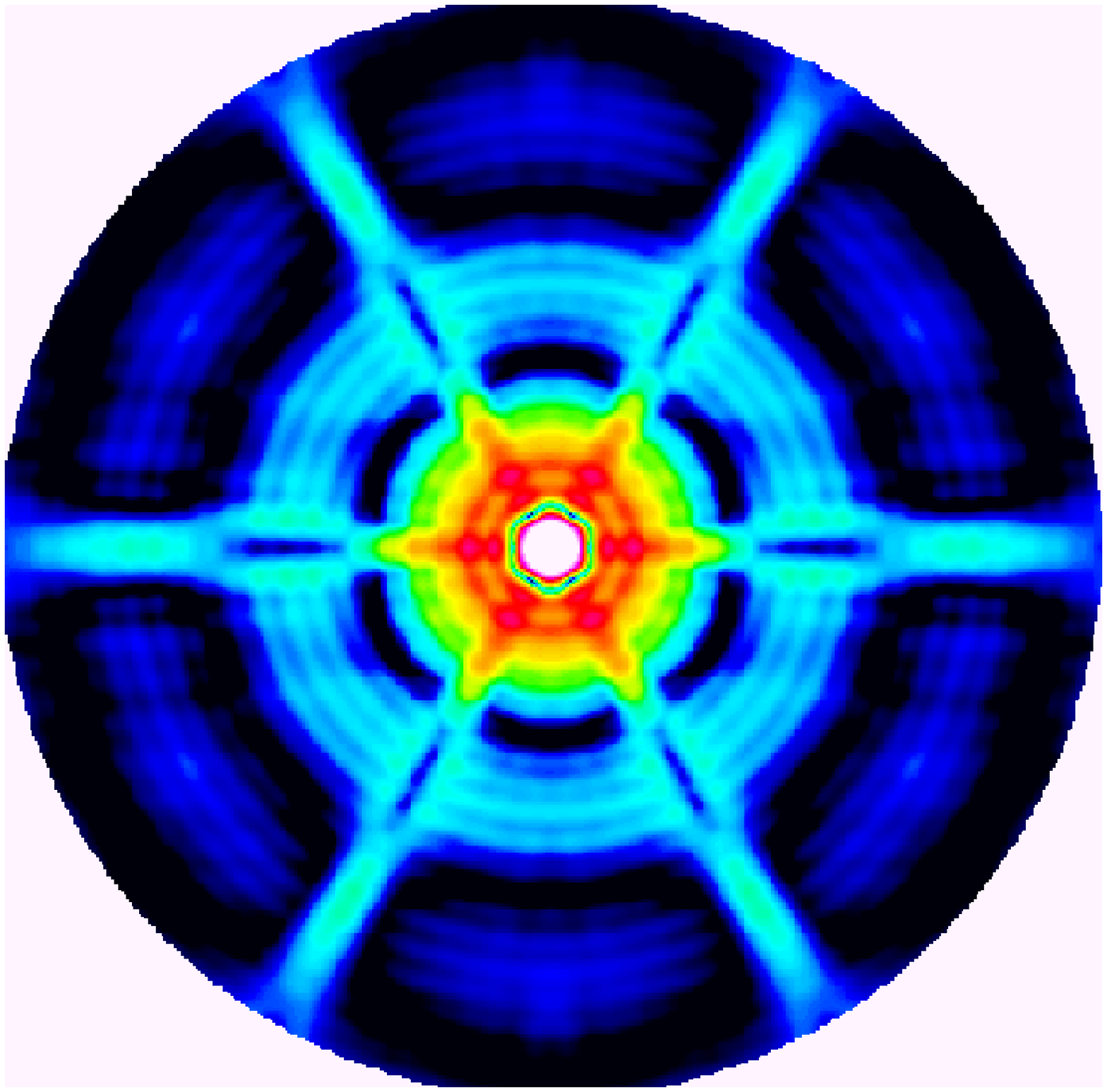}
   \includegraphics[angle=-90,width=6.cm]{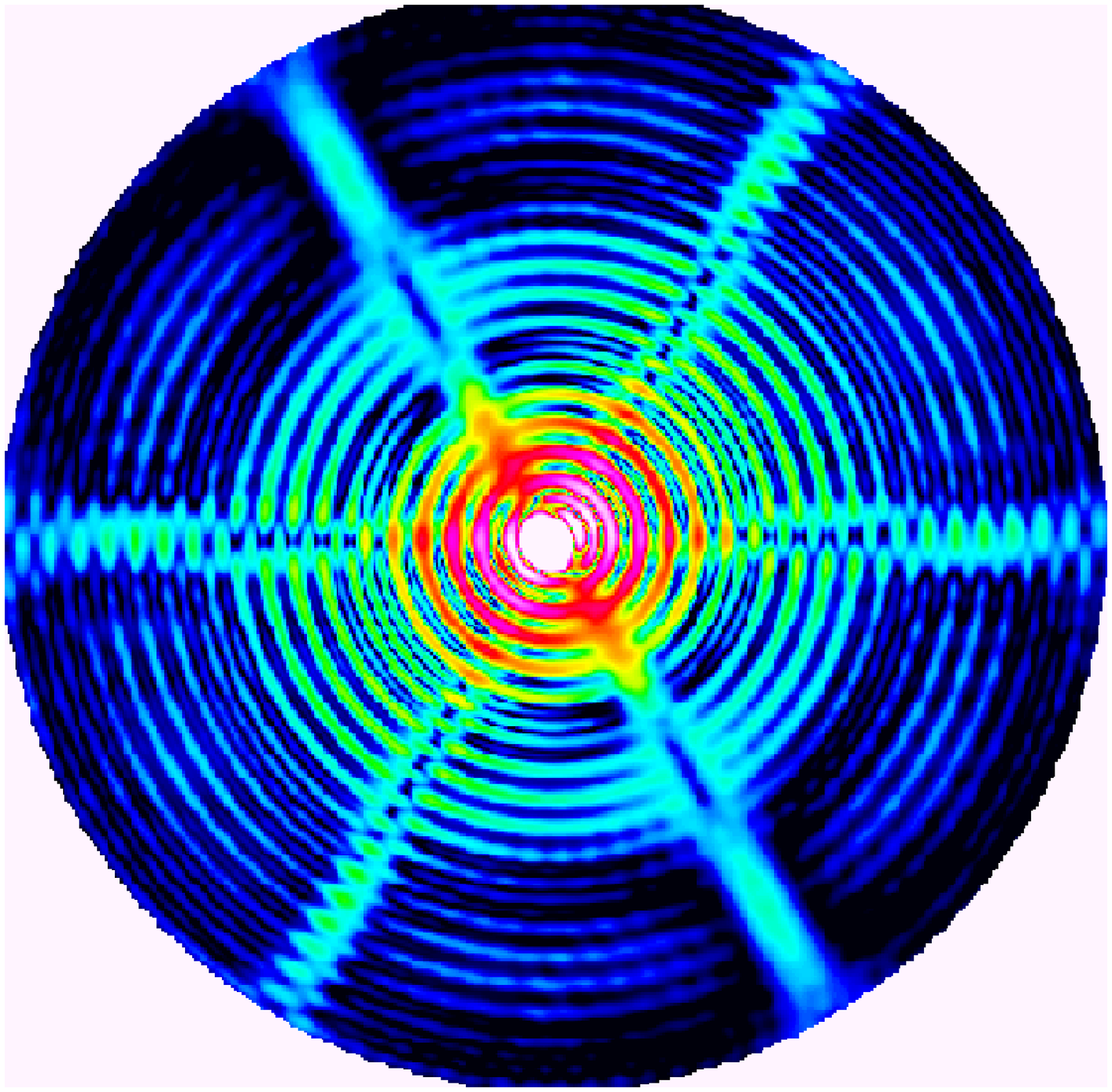}
   \includegraphics[angle=-90,width=6.cm]{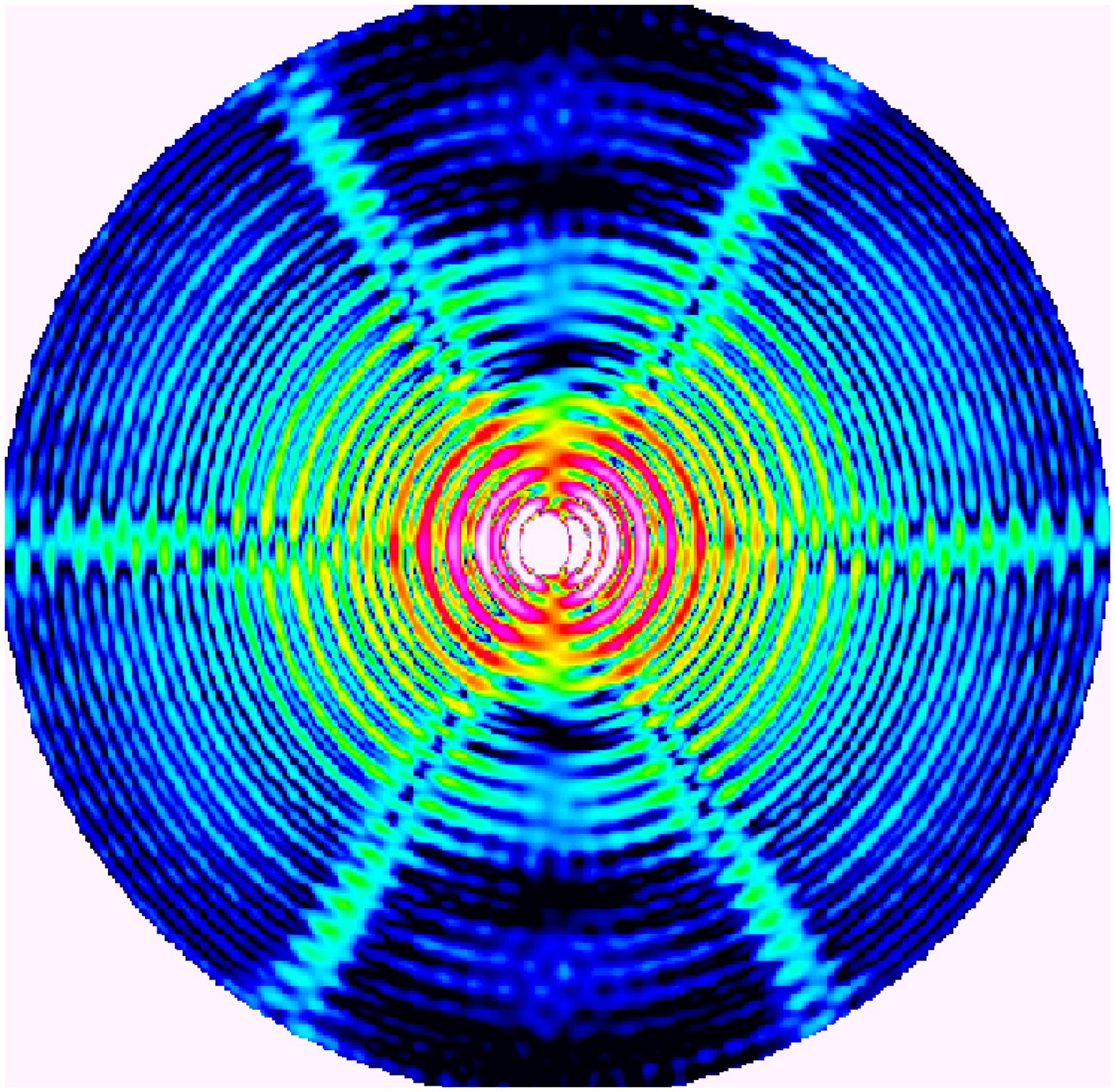}
   \caption{Feed configuration for the receiver (top) and antenna patterns 
     within a radius of 6\deg for beam 1, 2, and 8 (from top to bottom). The
     display is restricted to levels between -20dB (white) to -50db (black).}
   \label{Fig_Pattern}
\end{figure}

For the 13 beams we distinguished three characteristic cases, the central beam
and one feed from the inner and outer ring of feeds (Fig. \ref{Fig_Pattern}
top). The feed horns are equipped with two pairs of orthogonal-linear probes,
which, while in parallactic tracking, rotate relative to the feed support
legs. In turn the complex aperture distribution function changes with time. We
approximate this rotation with an accuracy of $\pm 5\deg$.  The fact that the
three feed support legs cause a global six-fold symmetry for the antenna
pattern simplifies the situation considerably. We need only to calculate 5
individual patterns for each of the rings with radial feed offsets of
29\farcm1 and 50\farcm8, respectively.

%-----------------------------------------------------------------------
%\begin{figure*}[!t]
%\end{figure*}
%-----------------------------------------------------------------------

Fig. \ref{Fig_Pattern} shows the feed configuration for the multibeam
system\footnote[1]{http://www.atnf.csiro.au/research/multibeam/.overview.html}
and displays the antenna response pattern within a radius of 6\deg~ for feeds
1, 2, and 8. We see characteristic changes due to the beam displacements. The
main beam develops a coma lobe that increases in strength with feed
offset. Sidelobe structures become asymmetric and have a more pronounced
ringing with increasing feed offset. An important feature of the multi-feed
system is that the azimuthal asymmetries in the antenna pattern due to the
feed offsets decrease on average with increasing distance from the main
beam. Extrapolating this property to far sidelobes, offset from the main beam
by tens of degrees, implies that differences between individual feeds should
vanish. Indeed we verified this behavior as described in Sect. \ref{farSL}.

In the correction algorithm it is not necessary to use a pattern with all the
details shown in Fig. \ref{Fig_Pattern} as the LAB data were measured on a
grid of only $0\fdg5 \times 0\fdg5$.  We have therefore modeled the inner
6\fdg0 of the antenna patterns on a grid containing 468 cells in cylindrical
coordinates. The cells have an azimuthal extent of 10\deg and a radial extent
of 0\fdg2 out to 2\fdg2 from the main beam, a radial extent of 0\fdg3 out to
2\fdg5, and of 0\fdg5 thereafter.

\subsection{The  far sidelobes}
\label{farSL}

The far sidelobes arise predominantly by reflections from the feed support
legs, causing the so-called stray-cones, and at the rim of the telescope, the
spillover. The location of these structures on the sky was determined from the
telescope geometry and verified by numerous tests made with a transmitter.
The sidelobe levels from the spillover lobes were estimated from the edge
taper of the receiver feed. For the stray-cones we used estimates from the the
aperture blocking by the feed legs. These estimates were also based on
sidelobe levels found previously for the telescopes at Effelsberg, Dwingeloo
and Villa Elisa \citep{Kalberla1980,Hartmann1996,Bajaja2005,Kalberla2005}.
The far sidelobe components were adjusted individually in a way similar to
that described by \citet{Kalberla2005}, searching for a consistent solution of
Eq.  \ref{EQdeconvolve}.

\subsection{Ground reflections}

Sidelobes that touch the ground receive broad-band thermal noise, but to first
order no line radiation is expected. However, depending on the ground cover,
reflections may happen. While scrub acts as an absorber, low grass and soil
can reflect 21-cm line radiation. We determine this contribution by surveying
the ground around the telescope, and then using ray-tracing methods to
estimate the effective sidelobe response. We used an albedo of 0.2.

\section{Removal of instrumental baselines}
\label{Baseline}

\subsection{Preliminary considerations}

In many cases, stray-radiation produces extended profile wings that mistakenly
may be interpreted as instrumental baseline structure. A correction for
stray-radiation could cause negative intensities if applied after the baseline
correction. It is therefore essential to apply baseline corrections only {\it
  after} the stray-radiation is eliminated. For this reason we had to deviate
from the processing described in \citetalias{Naomi2009}; for the first data
release no correction for stray-radiation was applied. We used {\it Livedata}
as described in their Sect. 2.3 for the bandpass correction, but without the
post processing for baselines according to Sect. 2.3.3 in
\citetalias{Naomi2009}. The algorithm described below is in lieu of the
post-processing described in \citetalias{Naomi2009}.

After scaling the reduced data to antenna temperatures corrected for
atmospheric absorption (Sect. \ref{Cal}) the stray-radiation was subtracted
and afterwards the instrumental baseline was corrected. Flags set by {\it
  Livedata} or our own software to indicate the presence of RFI were taken
into account, and affected channels were disregarded for all of the baseline
fits described below.

\subsection{The algorithm}

For each spectrum of the stray-radiation-corrected GASS database we used the
LAB survey for an initial estimate of the velocity extent of the expected \hi~
emission. After smoothing the interpolated LAB profile to an effective
resolution of $\Delta v_{1/2} = 5 $ \kms, we used velocities where $T_{LAB} <
0.9 $ K as the range for an initial fit of the instrumental GASS baseline. In
addition we folded the LAB profile in frequency to mimic the second spectrum
used by {\it Livedata} to determine the instrumental bandpass
\citepalias[][Fig. 3]{Naomi2009}. Flags were transferred accordingly.
  
GASS spectra from both polarizations were averaged and smoothed to $\Delta
v_{1/2} = 5 $ \kms~ for each of the individual 5 sec dumps. Using the initial
estimate for the emission-free part of the profile we determined the
instrumental baseline by fitting a 9$^{th}$ order polynomial.  Any channel
with signal exceeding four times the rms noise over the fitted region was
considered to contain possible \hi~ emission and was then excluded from the
baseline region. The baseline fit was repeated with an 11$^{th}$ order
polynomial, again searching for potential emission that needed to be excluded
from the fit. As a measure of the fit quality we determined the residual rms
scatter within the baseline region. For a few percent of the data we found
that a 9$^{th}$ order polynomial provided a better fit. Accordingly we used
this fit. Finally we took into account that weak profile wings with
intensities below the automatic $4 \sigma$ cut-off level could affect the fit
parameters. To avoid such a bias we excluded 6 more channels on each side of
the \hi~ emission from the baseline region that was fitted.

After these initial steps we considered both spectra for each polarization
individually. The spectra were smoothed and fitted with the parameters
determined previously to find out whether some more features needed to be
excluded from the baseline region. After that the final baseline fit was
applied to the original observations.

The algorithm and the parameters described above were tested and optimized by
varying the polynomial order and trying other functional forms.  We
particularly tried to avoid high order polynomials.  As described in
\citetalias{Naomi2009}, the average quotient spectrum for each of the scans
(visible as patches in Fig. \ref{Fig_Map2}, bottom) was fitted with a
$15^{th}$ order polynomial. Figure 3 of \citetalias{Naomi2009} demonstrates
that emission features are not affected by this fit. After this global
bandpass fit there remain some time dependent fluctuations that need to be
corrected independently for each spectrum with 5 sec integration time. A first
guess would be that a similar polynomial is needed to get rid of such
fluctuations. This is fortunately not the case. In the initial {\it Livedata}
reduction for the first data release a $10^{th}$ order fit for post-processing
of individual spectra was sufficient. Our results confirm this finding with
minor modifications for our procedure. We tested lower order
polynomials. These led for many regions to obvious baseline defects. We found
spurious features in emission free regions that could only be avoided by
allowing higher orders.

Assuming that some parts of the residual baseline ripples could be caused by
standing waves, we also tested sine wave fits in addition to
polynomials. Sinusoid solutions were found to be insignificant and were
discarded. We carefully checked whether errors in the LAB data might affect
the GASS baseline correction, but the iterative refinements that were applied
successfully avoided any biases. We tested also whether the high order
polynomial fits might have affected weak emission features. We found no
indications for such problems. In particular the RFI feature observed in March
2006, appearing similar to a HVC (Sect. \ref{mar06}), remained unaffected by
the fit. Also testing whether profile wings in data that remained uncorrected
for stray-radiation could be removed by baseline fitting did not indicate an
unwanted overcorrection by the baseline fit (see Sect. \ref{Results}).

Some residual systematical baseline problems remain in a few regions, on
average typically at a level of $\la 40$ mK but occasionally up to 100
mK. This is close to the rms uncertainties of $\sim 60$ mK for individual maps
as discussed in Sect. \ref{Results}. We were unable to identify the reason for
such systematic baseline errors. They may be caused by RFI or bad fits but
also by uncertainties in the stray-radiation correction.

\section{Calibration}
\label{Cal}

%-----------------------------------------------------------------------
\begin{figure*}[!t]
   \centering
   \includegraphics[angle=-90,width=9cm]{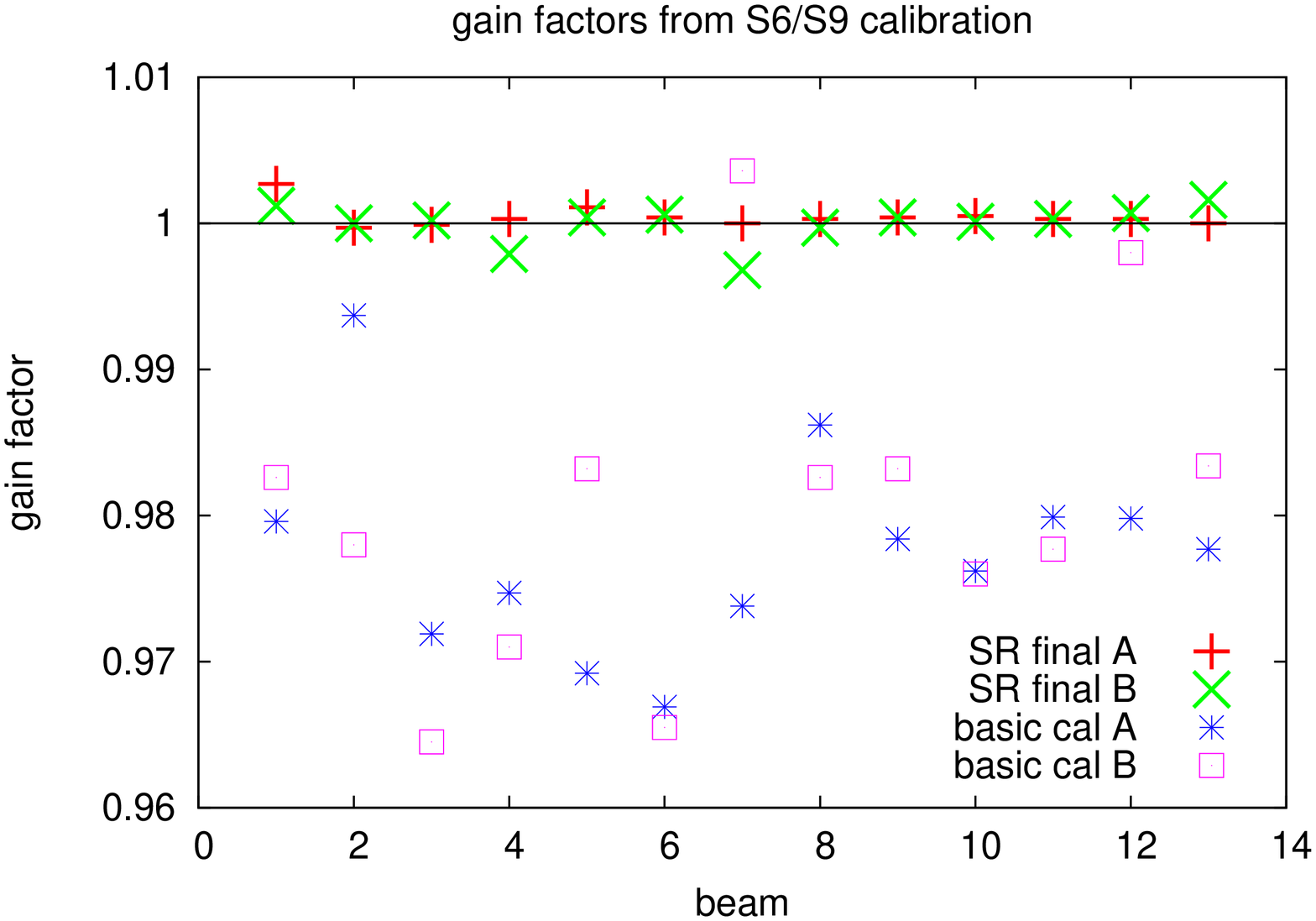}
   \includegraphics[angle=-90,width=9cm]{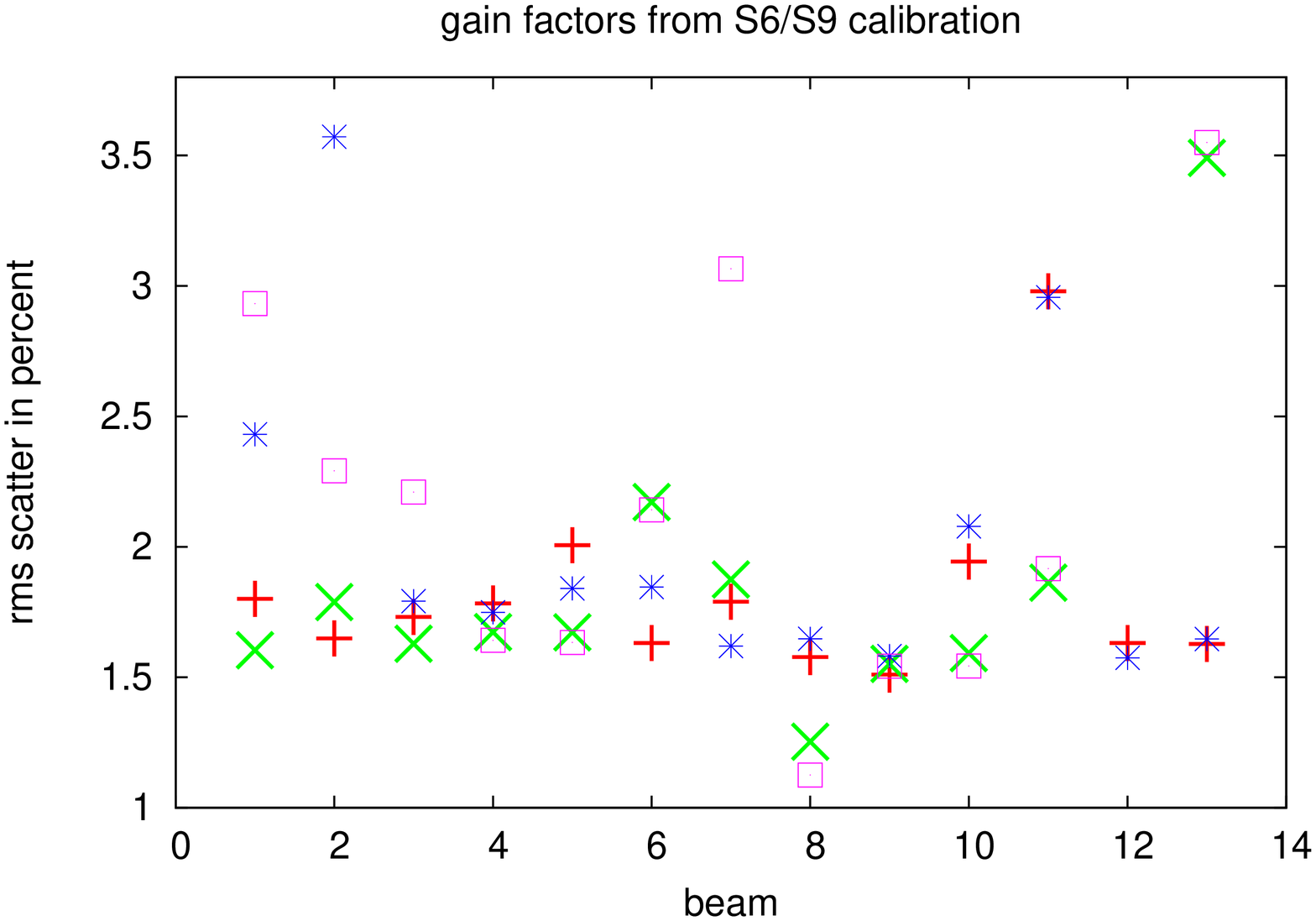}
   \caption{Left: Gain factors applied to the initial calibration (blue stars
     and purple squares) and cross check after the final data reduction (red
     plus signs and green X's). Right: RMS scatter in gain calibration for
     individual feeds, before and after correction for stray-radiation (same
     symbols as in left panel).}
   \label{Fig_Calib}
\end{figure*}
%-----------------------------------------------------------------------

An accurate correction for stray-radiation demands a careful
calibration. Errors in the antenna temperature scale cause a mismatch with the
stray-radiation correction term in Eq. \ref{EQdeconvolve}. The sidelobe
response is frequency dependent; calibration errors, therefore, cause not only
an increased scale error but also affect the profile shape.  We used regular
observations at the standard positions S8, S9, and S6 \citep{Williams1973} and
determined average calibration factors for each of the observing sessions. An
initial temporary calibration was determined from uncorrected data
\citepalias{Naomi2009}. After correcting the data for stray-radiation we
redetermined the $T_B$ calibration factors and applied these new factors for
the following calculations.

Table \ref{table2} lists the final calibration data; the brightness
temperature scale is matched to a common temperature scale first established
by \citet{Williams1973}. The brightness temperature for most of the major
surveys are matched through this reference; this includes the Hat Creek
surveys \citep{Weaver1973,Heiles1974}, the Leiden-Green Bank survey
\citep{Burton1985}, southern sky surveys by \citet{Kerr1981}, the Bell Labs
survey \citep{Stark1992}, the LAB survey, and Effelsberg and Arecibo data
\citep{Peek2008,Winkel2010}.

A few calibration regions have been mapped by \citet{Kalberla1982}. They
calculated the beam response in detail and provided also calibration factors
as a function of the beam shape. We use factors appropriate for a 14.4\arcmin
beam.

The easiest way to determine the calibration factors is to use peak
temperatures, but line integrals over a limited velocity range are more
accurate \citep{Williams1973} and we preferred this method. Please note that
velocities listed in Table \ref{table2} are integration limits while
velocities given by \citet{Williams1973} are center velocities for the edge
channels \citep[see][]{Kalberla1982}. Figure \ref{Fig_Calib} (left) compares
overall gain factors applied to the initial calibrations (blue) with those
after completing the final corrections (red and green). The overall
calibration uncertainties (right) for most of the feeds could be reduced to
typically 1.7\%. Only two receivers were less stable with an rms scatter of
3\% and 3.5\%. The uncertainties are of statistical nature. Combining
observations with several beams at different seasons we expect average
temperature scale uncertainties well below 1\% in the final data.  We note
that interference can cause a loss of calibration in some spectra that would
be difficult to detect as it would appear as a scale factor over an entire
spectrum.  We believe that this effect is small in the overall survey but may
be present in some final spectra.

\begin{table}
\caption[]{$T_B$ calibration positions and parameters}
\begin{tabular}{lccccc}
\hline
Pos. & l  & b  & Peak T$_{B}$  &  Area & Vel. range  \\
& {} & {} & {(K)} & {(K~\kms)}  & {(\kms)} \\
\hline
%\startdata
S8 &  207\fdg00 & --15\fdg00 &75.6 & 879.5 & --5.1 to +22.3 \\
S9 &  356\fdg00 & --4\fdg00 &83.5 & 930.8 & --1.5 to +15.2 \\
S6 &  1\fdg91 & 41\fdg42 &51.0 & 285.5 & --5.8 to +4.8 \\
%\enddata
%\end{flushleft}
\label{table2}
\end{tabular}
\end{table}

\section{RFI mitigation}
\label{RFI}

\subsection{Replacement of data flagged by {\it Livedata} }
\label{livedataflagging}
 
\begin{figure*}[!t]
   \centering
   \includegraphics[angle=-90,width=9.cm]{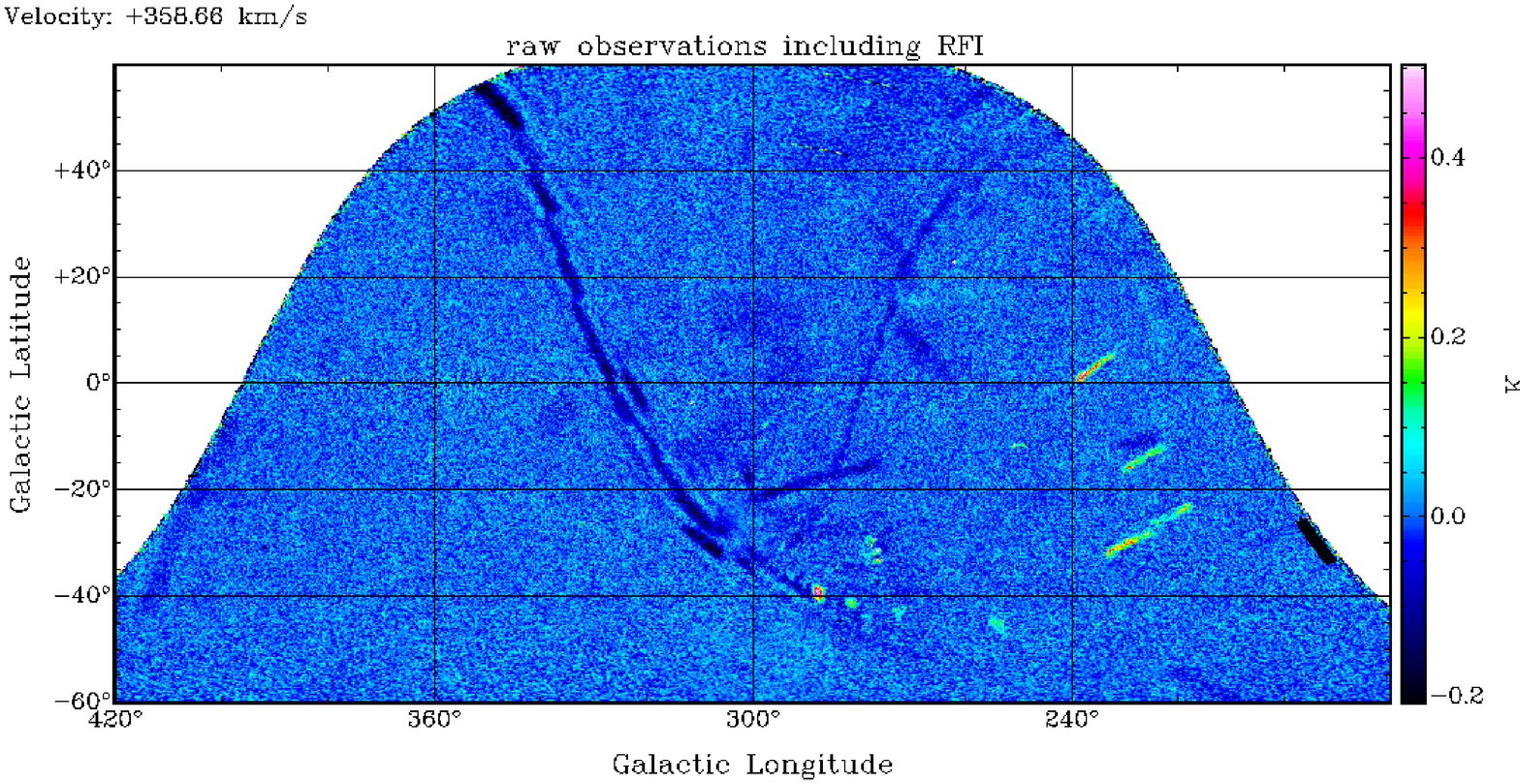}
   \includegraphics[angle=-90,width=9.cm]{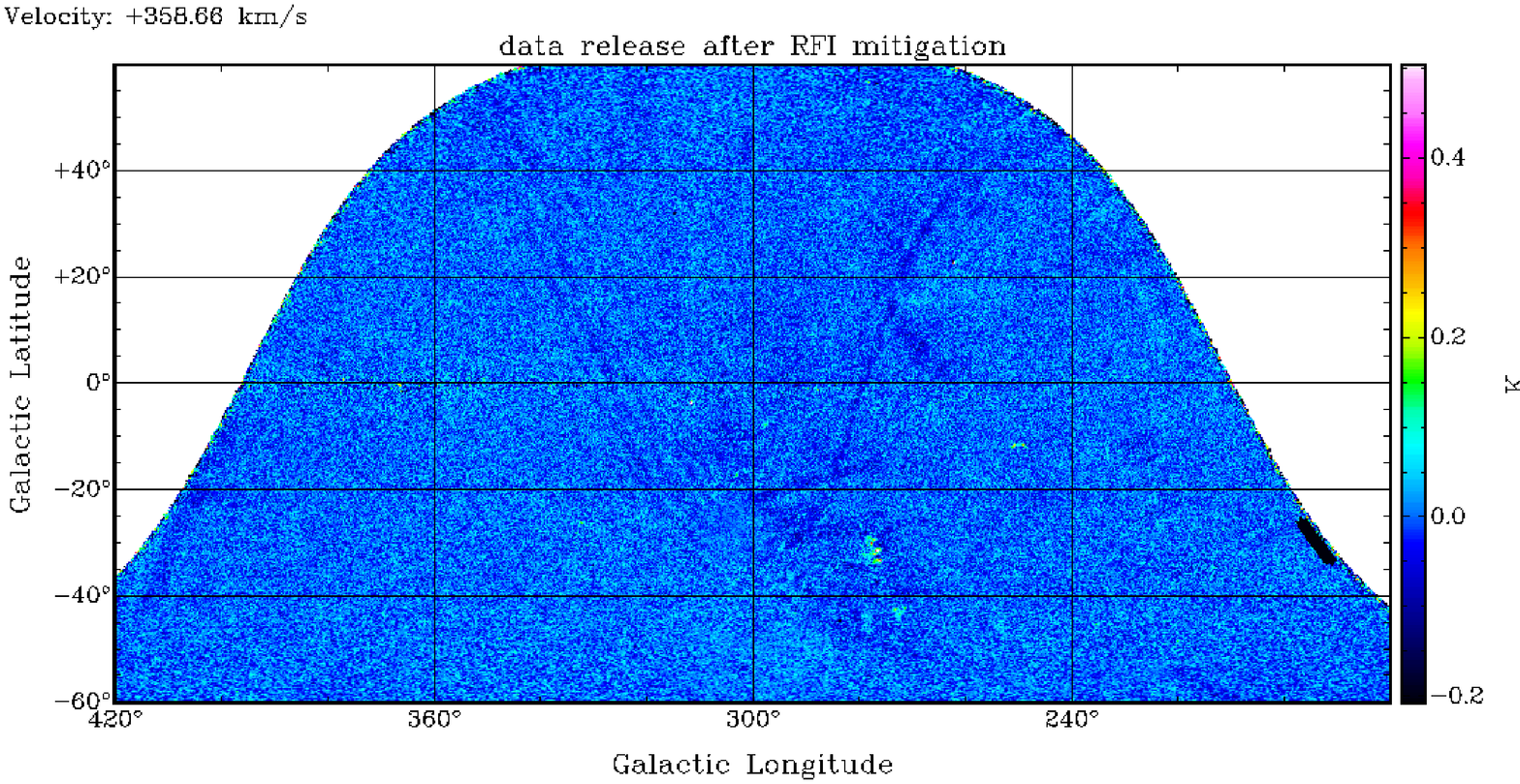}
   \caption{Typical example for extended interference patterns that were
     flagged by {\it Livedata} in the first stage of the data reduction. Left:
     as observed, right: after RFI removal using LAB data as described in
     Sect. \ref{livedataflagging}. The intensity scale is linear for $ -0.2 <
     T < 0.5$ K. Some weak structures remain but are mostly at $|T| < 40$ mK }
   \label{Fig_RFI1}
\end{figure*}

Narrow-line interference was flagged during the basic data reduction with {\it
  Livedata} \citepalias[][Sect. 2.3.2]{Naomi2009}. Most of this type of RFI
occurs at a fixed topocentric frequency, so as the telescope scans across the
sky the RFI moves in successive channels in the LSRK frame. We use the
kinematical (radio) standard of rest definition with a velocity of 20 \kms~ in
direction ra,dec = [270\deg,+30\deg] (epoch 1900.0).  Most of the features are
observed to stretch over large angular distances and affect fields that have
been observed independently at different seasons (Fig. \ref{Fig_RFI1}).  This
behavior implies that the RFI remained approximately constant for long periods
in time, affecting most of the data.

Profiles with more then 30 flagged channels were completely discarded during
imaging; for less serious contaminations the flagged data were replaced. After
correcting profiles for stray-radiation and instrumental baseline, we
interpolated LAB survey data to replace the flagged channels.  Alternatively
we have tried to replace data by interpolating the GASS spectra using the
nearest ten good data points.  Using LAB data resulted in all cases in much
cleaner maps. Still, some weak stripy features remain that appear to move
across the sky in successive velocity channels but mostly these artifacts
remain within the noise (Fig. \ref{Fig_RFI1}). About 0.1\% of all data have
been flagged by {\it Livedata} and were replaced by interpolated data in this
first step. We kept the {\it Livedata} flags to avoid such interpolated data
getting propagated by the RFI mitigation described in the following sections.

\subsection{Median filter RFI rejection}
\label{median}
%-----------------------------------------------------------------------
%-----------------------------------------------------------------------

\begin{figure*}[!t]
   \centering
   \includegraphics[angle=-90,width=9.cm]{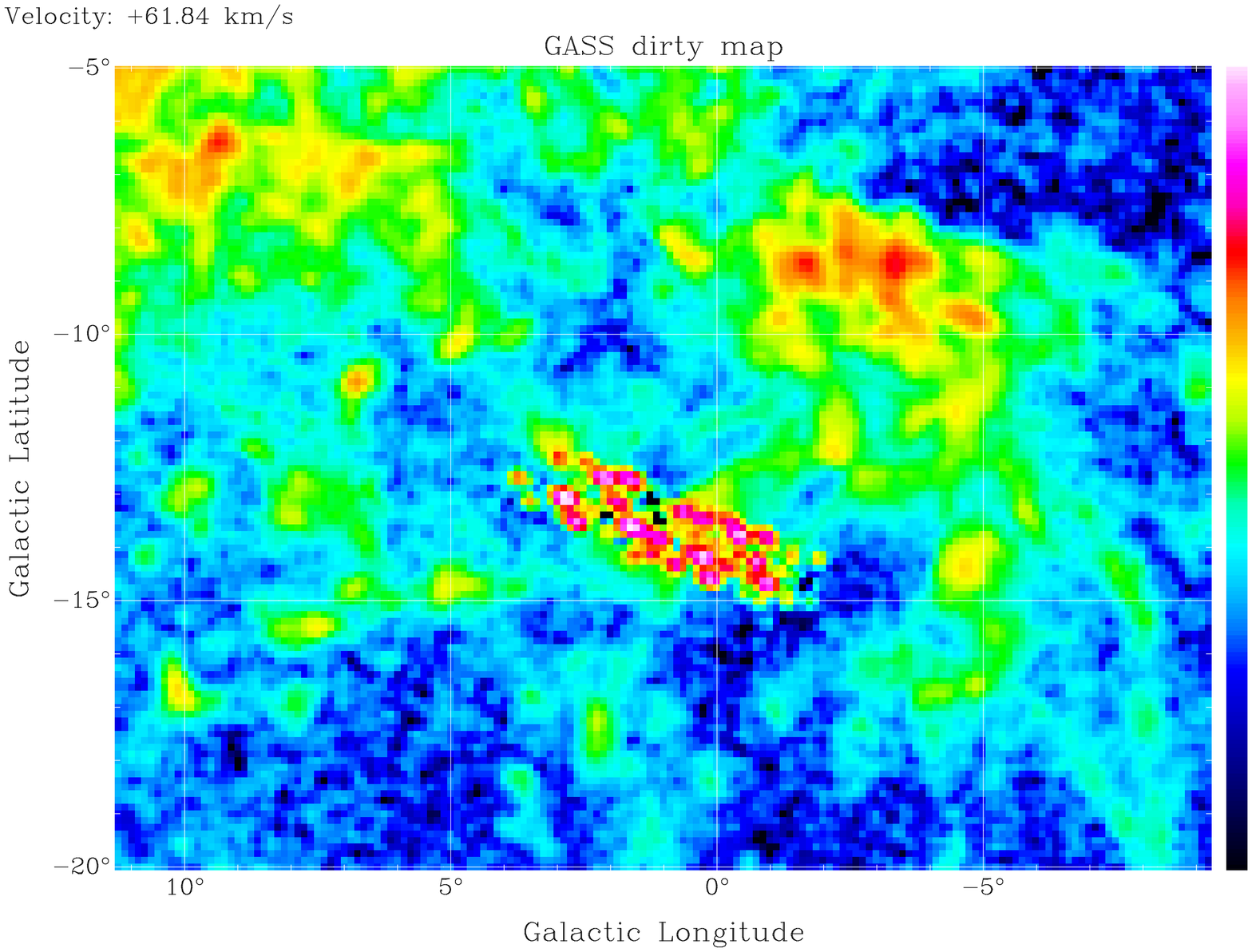}
   \includegraphics[angle=-90,width=9.cm]{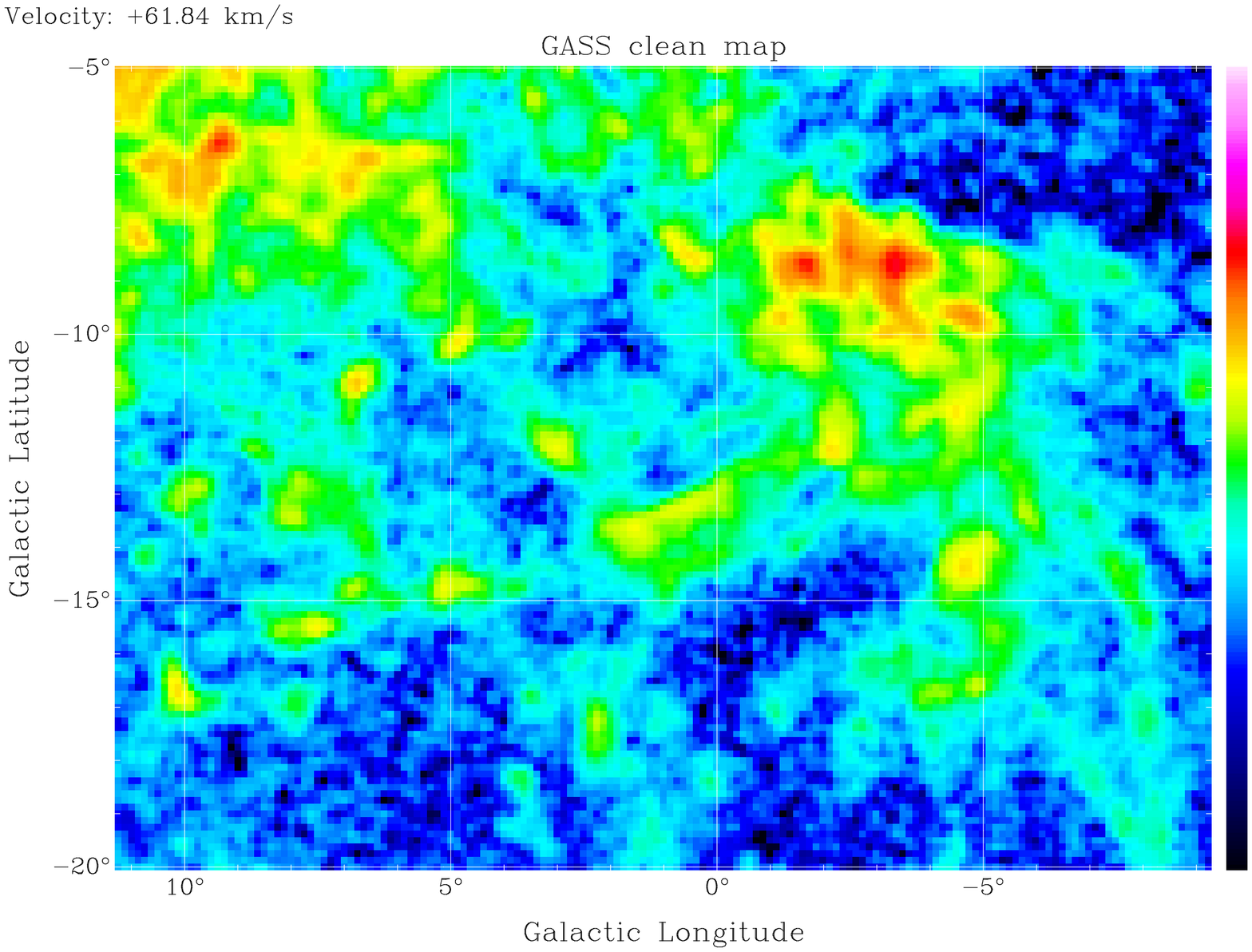}
   \caption{HI emission at ${\rm v_{lsr}} = 61.8 $ \kms.  Data heavily
     affected by RFI in scanning direction of the telescope (left) are
     compared with the clean map (right). The intensity scale for the range $
     -.04 < T < 20$ K is logarithmic, emphasizing low level emission. Yellow
     colors correspond to a brightness temperature of $\sim 1$ K.}
   \label{Fig_RFI}
\end{figure*}

Even after removal of the channels flagged in the initial data reduction,
numerous instances of RFI remain in the data. Some appear as point sources,
but the most obvious show ``footprints'' of the 13-beam system that move on
the sky in scanning direction (Fig.\ref{Fig_RFI}). As before, the motion in
successive velocity channels is caused by changing LSR projections of the
fixed Topocentric RFI frequencies.  Features of this kind were found
previously in the HIPASS and are described in detail by
\citet[][Fig. 7]{Barnes2001}. For the first data release the remaining RFI was
removed by median gridding when calculating FITS images
\citepalias[][Sect. 2.3.5]{Naomi2009}.

Here we follow a different strategy, taking advantage of the fact that in GASS
typically 40 individual spectra contribute to the final data in every
resolution element.  First, for any observed position in the sky we selected
all observed profiles within an angle $\epsilon$ of typically a few
arc-minutes. From these data we calculate the mean for each channel, the
corresponding standard deviation, and the median, disregarding data flagged
either by {\it Livedata} or our own post-processing. The average rms noise
level $T_{RMS} = \sigma (T)$ and its one sigma deviation $\sigma (T_{RMS})$
from the average was determined for the low level emission part of the profile
($T_B < 0.5$K).

We have two tests for RFI.  In the first, channels with rms deviations $\Delta
T_{RMS} > 3 \sigma (T_{RMS})$ were considered as highly suspicious for RFI
contamination and investigated in more detail. We use as a second indication
of RFI circumstances where the mean and median differ from each other by more
than one standard deviation.

To eliminate outliers for the channels that were selected in both of the two
tests we determined if individually observed brightness temperatures deviate
from the median by more than two times the standard deviation. In the case of
a purely random distribution this would exclude 5\% of the good data, and we
consider this an acceptable cost to eliminate true RFI.  We flag outliers and
repeat, calculating new mean and median estimators and once more excluding any
outliers.

Our two sigma limit may be compared with the more rigorous method for the
elimination of suspect data proposed by \citet{Peirce1852} which is based on
probability theory. Accordingly a rejection of a data point that deviates from
the mean at a two sigma level is justified if there are 13 independent
observed data values. For $\epsilon = 6\arcmin$ (see below) we have on average
40 individual profiles and a two sigma limit is appropriate if about 10\% of
the data are suspect. This condition is valid for many cases but we find also
situations where 20\% of the measurements are outliers. In this case a cutoff
at $1.6~ \sigma(T_{RMS})$ would be more appropriate according to
\citet{Peirce1852}. We deviate from Peirce's criterion in two ways; to
simplify the calculations we use a fixed $2~ \sigma(T_{RMS})$ cutoff and we
define outliers by their deviations from the median and not, as originally
proposed by Peirce, from the mean. The latter is essential since outliers
caused by RFI can be found at ten to hundred times the standard deviation, a
circumstance that was not considered by Peirce; in such cases the mean is
usually seriously affected.

To replace values outside the two sigma range, the best available estimate is
the median value scaled by the proper beam weighting function, for details we
refer to the extended discussion by \citet[][Sect. 4.2.2]{Barnes2001}. We
replace the previously selected data points by the weighted median and set a
flag to distinguish medians from observed values. A fraction $\sim 5 \cdot
10^{-4}$ of the data is affected by RFI and replaced by the weighted median.

One of the criteria that we used to detect RFI was the fact that mean and
median should not differ by more than one standard deviation. We applied this
criterion also to test whether the median estimator is consistent with the
mean of all data after exclusion of the outliers. Both methods were found to
produce nearly identical results. An example of RFI in scanning direction and
the result after application of our median filer is given in
Fig. \ref{Fig_RFI}.

\begin{figure*}[!t]
   \centering
   \includegraphics[angle=-90,width=9.cm]{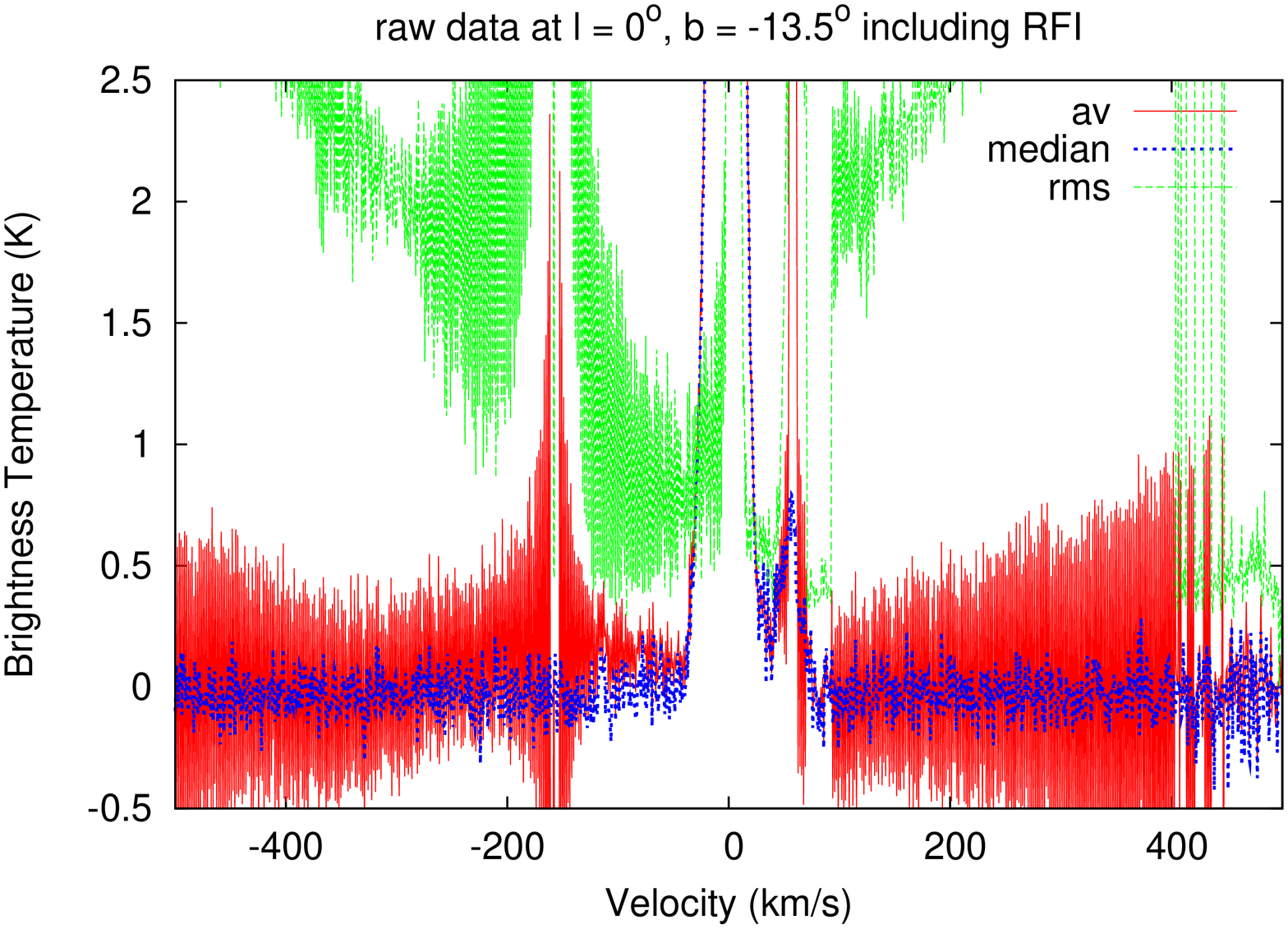}
   \includegraphics[angle=-90,width=9.cm]{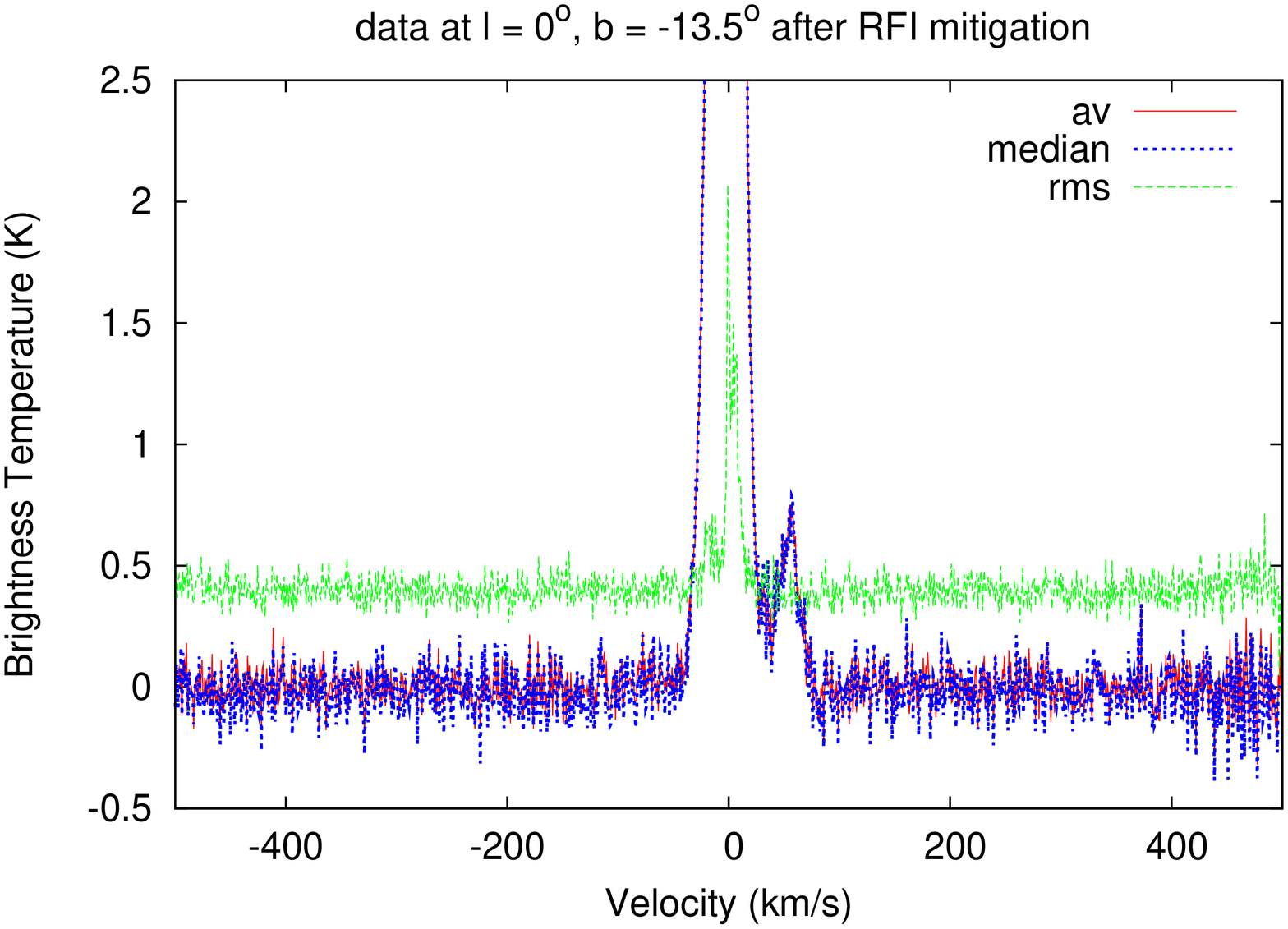}
   \caption{Data at the position $l = 0\deg$, $b = -13\fdg5$ before and after
     RFI mitigation. The average is plotted in red, the rms scatter in green,
     and the median in blue. }
   \label{Fig_RFIprof}
\end{figure*}

Figure \ref{Fig_RFIprof} shows spectral details for one position in this
field. Displayed are the mean (red), the rms scatter (green), and the median
(blue) derived for all profiles that are closer than $\epsilon < 6\arcmin$ to
the position $l = 0\deg$, $b = -13\fdg5$. Data in the left plot suffer from
strong interference at ${\rm v_{lsr}} \sim 157$ \kms, $55 \la {\rm v_{lsr}}
\la 58$ \kms, and at $ {\rm v_{lsr}} \ga 400$ \kms. Only a part of the data
was flagged by {\it Livedata} and replaced by values from the LAB. Despite all
defects the median (blue) is essentially unaffected by RFI. The data after RFI
elimination are shown in the plot at the right side. Profiles for mean and
median are within the noise identical, the rms is close to the expected value
of 0.4 K and shows no strong enhancements. We find some fluctuations close to
the line emission at ${\rm v_{lsr}} \sim 0$ \kms. This may have been caused by
genuine fluctuations in the \hi~ line emission. To avoid a possible
degradation of such fluctuations no automatic RFI cleaning was applied for
$T_B \ga 0.5$K. These two plots demonstrate how greatly the rms is affected by
RFI, and how deviations between median and mean are obvious. Our RFI filter is
triggered by such deviations.

\subsection{The beam-specific parameter $\epsilon$}

A critical parameter for RFI filtering is the radius $\epsilon$ for a
comparison of the data at adjacent positions. A large value provides a large
statistical sample but at the same time point sources or steep gradients in
the brightness distribution may be affected by the filter. We found in tests
that $\epsilon = 6\arcmin$ is sufficiently robust.  Tests with \hi~
point-sources have shown that larger values could degrade the source
distribution. Our result is in excellent agreement with
\citet{Barnes2001}, and our median estimator is
identical with the parameter $r_{max} = 6\arcmin$ that was used for HIPASS.  

%-----------------------------------------------------------------------
%-----------------------------------------------------------------------

%-----------------------------------------------------------------------
%-----------------------------------------------------------------------

\subsection{RFI  in emission line regions}

RFI that falls in channels containing \hi~ emission was usually removed
successfully by the above procedure, except in regions with strong spectral
gradients. Examples of such critical cases are positions with strong
absorption lines. Sources with a continuum flux exceeding 200 mJy were
excluded from RFI rejection.

To avoid any possible degradation of the signal due to an automatic RFI
filtering, the median filters described in Sect. \ref{median} were restricted
to data with low brightness temperatures. We used a limit of $T_{max} \la
0.5$K. The remaining data were checked visually. Only a few cases showed some
residual RFI. For these we increased the filter limits such that only channels
with sufficient large rms deviation $\Delta T_{RMS}$ were affected. We
verified that the remaining data were unaffected by the median filter. As a
general rule we tried to avoid as far as possible any modification of the \hi~
emission lines through the RFI filtering process. This implies, however, that
some RFI may remain undetected and be present in the final data.

Some of the spectra were affected so severely by RFI that they needed to be
removed completely. We eliminated all profiles whose mean rms within the
baseline region exceeded the average noise level at its position by a factor
of three. About 0.3\% of the observed profiles were rejected for this reason.

\subsection{RFI in March 2006 }
\label{mar06}

For a short period between 18 and 21 March 2006 observations were degraded by
broadband RFI with quite different characteristics than any of the narrow band
RFI described above. The weak spurious lines had approximately Gaussian shapes
with center velocities in the range 270 to 310 \kms~ and velocity dispersions
of about 15 \kms. None of the RFI rejection methods described above applied to
this case.  These spurious lines were identified by fitting Gaussians; all
affected channels were then flagged. Baseline corrections were repeated,
finally all flagged data were replaced by medians as described
before. This way we recovered data that were discarded in the first data
release.

\subsection{Removal of bandpass ghosts}
\label{ghosts}

To maximize sensitivity and recover all the extended \hi~ emission, the GASS
was made using in-band frequency-switching with an offset of 3.125 MHz
corresponding to 660 \kms. The {\it Livedata} bandpass correction causes for
every real emission line feature an associated, spurious, negative image
(``ghost'') in the other band, displaced by $\pm 660$ \kms. Most of these
negative images fall outside the velocity coverage of the survey but high
velocity clouds (HVCs) with emission lines at $|v_{\rm lsr}| \ga 160 $ \kms\
cause ghosts. In particular, the Magellanic System causes a strong and
extended bandpass ghost at negative velocities.

As Fig. 3 of \citetalias{Naomi2009} shows, ghosts appear only in one of the
two IFs; the other band is unaffected. It is possible to produce maps without
ghost features if one uses only one IF for $|v_{\rm lsr}| \ga 160 $ \kms. A
second possibility is to flag ghosts and eliminate them with a treatment
analogous to one of the the RFI mitigation strategies discussed above.
Accordingly the ghosts were replaced by medians determined from the alternate
ghost-free IF band. Such maps have the advantage that they provide a better
sensitivity for most of the field but remaining biases can not be excluded.
We determined for both bands all HVC emission features with $|v_{\rm lsr}| \ga
160 $ \kms~ and $T_B > 0.2$ K (corresponding to $3 \sigma (T_{RMS})$ for
$\epsilon = 6\arcmin$) and flagged ghosts in the other band. These features
were treated analogously to the broadband RFI discussed in the previous
section.

\section{Imaging}
\label{Images}

For HIPASS, as well as for the first data release of the GASS, imaging was
performed by {\it Gridzilla} using weighted median estimation which is robust
against RFI and other artifacts. For the second data release, we use separate
and independent procedures for RFI rejection and imaging.  As described in the
previous section, our median filter is largely consistent with the HIPASS
median filter algorithm for a radius $\epsilon = 6\arcmin$. However, the major
difference is that we apply this filter only to individual channels if two
conditions apply: 1) an exceedingly large scatter is observed with a 3 times
larger dispersion than the mean dispersion caused by thermal noise and 2) mean
and median differ by more than one standard deviation.  Only data that deviate
from the median by $\Delta T_{RMS} > 2 \cdot \sigma (T_{RMS})$ are replaced by
the median estimate.

%-----------------------------------------------------------------------
\begin{figure}[!t]
   \centering
   \includegraphics[angle=-90,width=8cm]{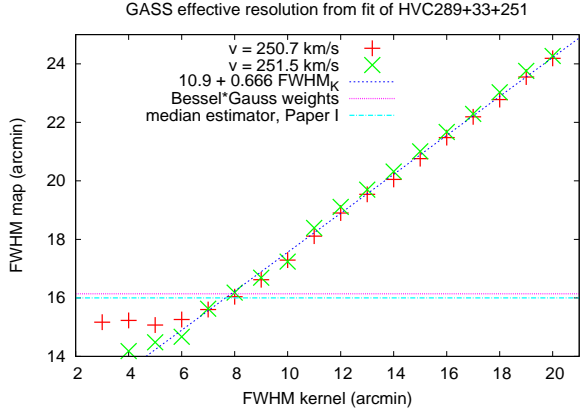}
   \caption{Effective FWHM resolution $W$ of GASS images as a function of the
     user-definable Gaussian interpolation kernel $W_k$. For of $W_k >
     7\arcmin$ we find $W \sim 10\arcmin9 + 0.666 W_k$. For comparison we show
     the effective resolution obtained by median gridding and by Bessel
     interpolation (horizontal lines). }
   \label{Gass_res}
\end{figure}
%-----------------------------------------------------------------------

%-----------------------------------------------------------------------
\begin{figure}[!t]
   \centering
   \includegraphics[angle=-90,width=8cm]{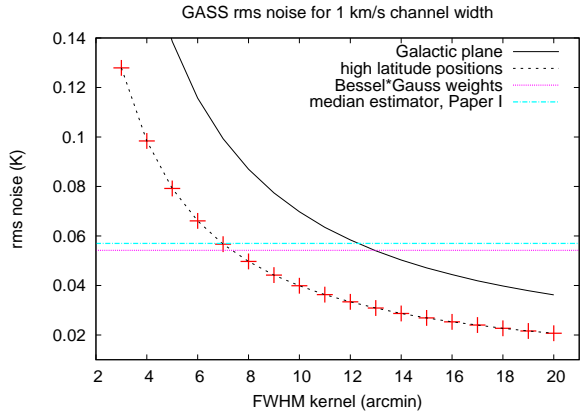}
   \caption{Effective rms uncertainties of GASS channel maps with 1 \kms~
     velocity resolution as a function of the user-definable Gaussian
     interpolation kernel. Crosses indicate fitted values determined at high
     Galactic latitudes, the solid black line represents upper limits of the
     noise, valid for most of the positions in the Galactic plane. For
     comparison we show the effective noise at high latitudes obtained by
     median gridding and by Bessel interpolation (horizontal lines).  }
   \label{Gass_noise}
\end{figure}
%-----------------------------------------------------------------------

For the data release described here the gridding was performed using a
two-dimensional Gaussian with a user-definable dispersion. For computational
reasons we disregard data outside a three sigma cutoff radius. The
interpolated value for that pixel is the weighted mean over all available
input data. Except for the irregular distribution on the sky of the input
data, this interpolation is equivalent to a linear convolution of the observed
brightness temperature distribution with the kernel function of the
gridder. The resulting effective beam is approximately Gaussian. For a source
width $W_s$, an effective average telescope beam width of $ W_t$, and a width
$W_k$ of the user defined kernel we expect a resulting FWHM width
\begin{equation}
  W = \sqrt{W_s^2 + W_t^2 + W_k^2} 
\end{equation}

We tested this relation by generating a series of images of the ultra-compact
high-velocity cloud HVC289+33+251 \citep{Putman2002,Bruens2004}. We used a
grid spacing of $4\farcm8$, corresponding to 1/3 of the FWHM beam size. This
object is isolated and has a FWHM diameter of $W_s = 4\farcm4$. Since it is
unresolved by the Parkes telescope at 1.4 GHz the effective beam $W$ can
easily be fitted. This was done for 18 FITS cubes with $3\arcmin < W_k <
20\arcmin$ at two velocities, $v = 250.7$ \kms~ and $v = 251.5 $ \kms. We
determined first an effective telescope beam width $W_t = \sqrt{W_k^2 - W_s^2}
= 14\farcm2 \pm 1\farcm0$ from the known width of kernel and source. This
value agrees well with the expected average FWHM beam width. The individual
beams have $14\arcmin < {\rm FWHM} < 14\farcm5$, with a mean of $ W_{av} =
14\farcm4$.

Next we deconvolve the observed and fitted FWHM width $W_{obs}$ of the image
for the finite width $W_s = 4\farcm4$ of the source, $W = \sqrt{W_{obs}^2 -
  W_s^2}$. In Fig. \ref{Gass_res} we plot $W$ as a function of the convolving
kernel width $ W_k$. To a good approximation we find for $ W_k \ga 6\arcmin$
the simple relation $ W = 10\farcm9 + 0.666 \cdot W_k$. For a kernel $ W_k =
8\arcmin$ we obtain $ W = 16\arcmin$, identical with the resolution of the
images calculated from weighted median estimators as presented in
\citetalias{Naomi2009}.

%-----------------------------------------------------------------------
\begin{figure*}[!t]
   \centering
   \includegraphics[angle=-90,width=14.5cm]{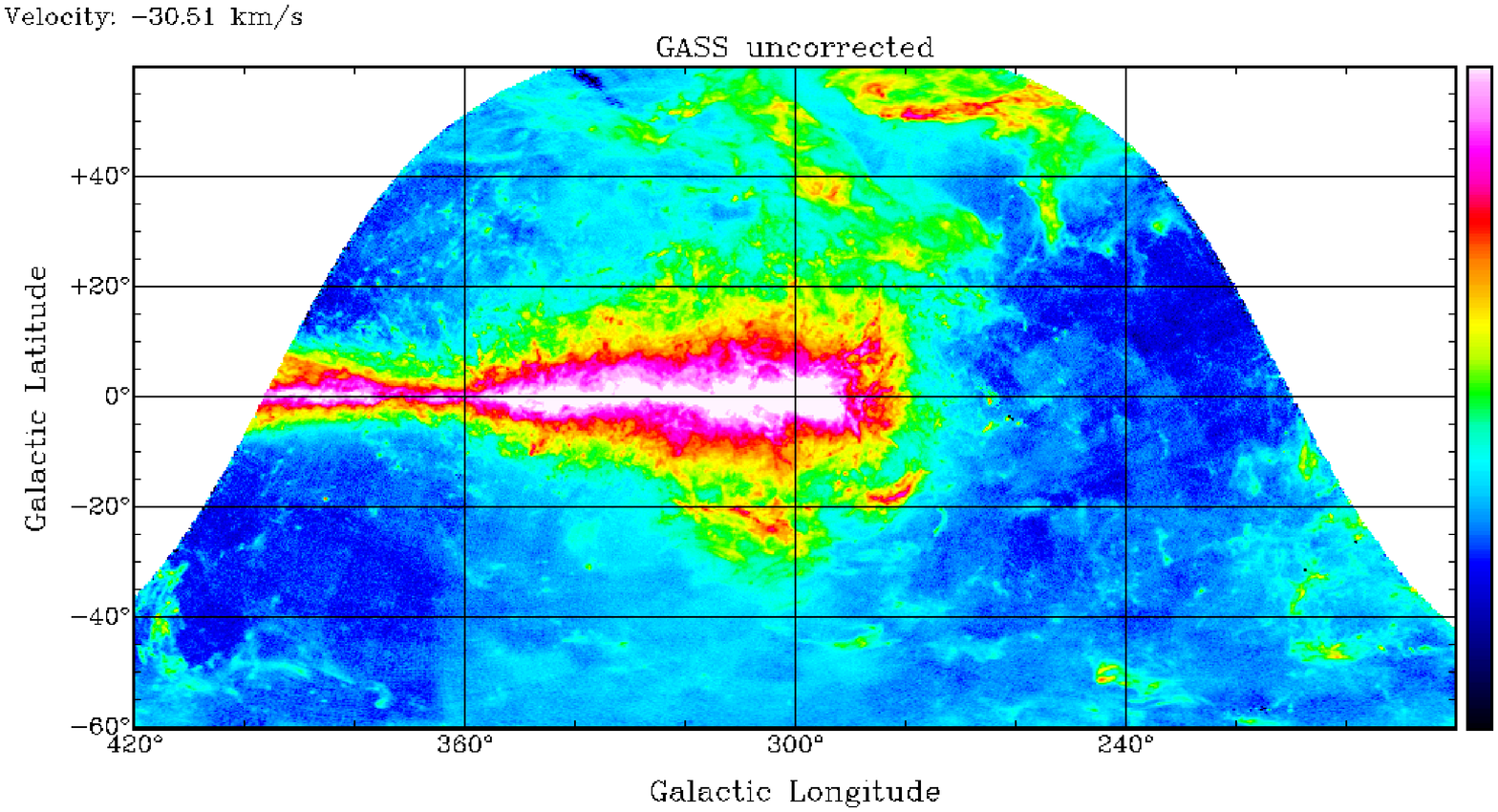}
   \includegraphics[angle=-90,width=14.5cm]{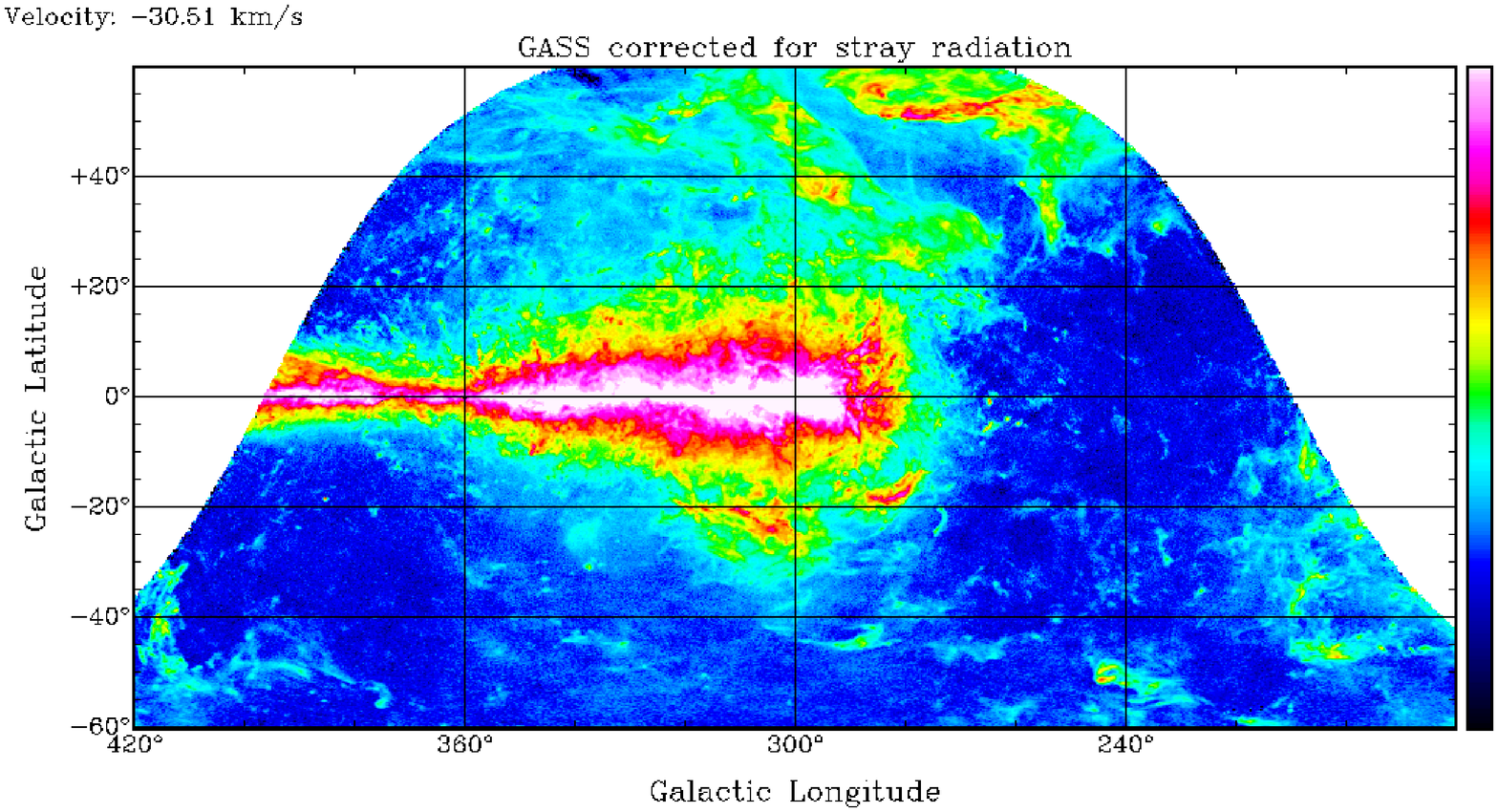}
   \includegraphics[angle=-90,width=14.5cm]{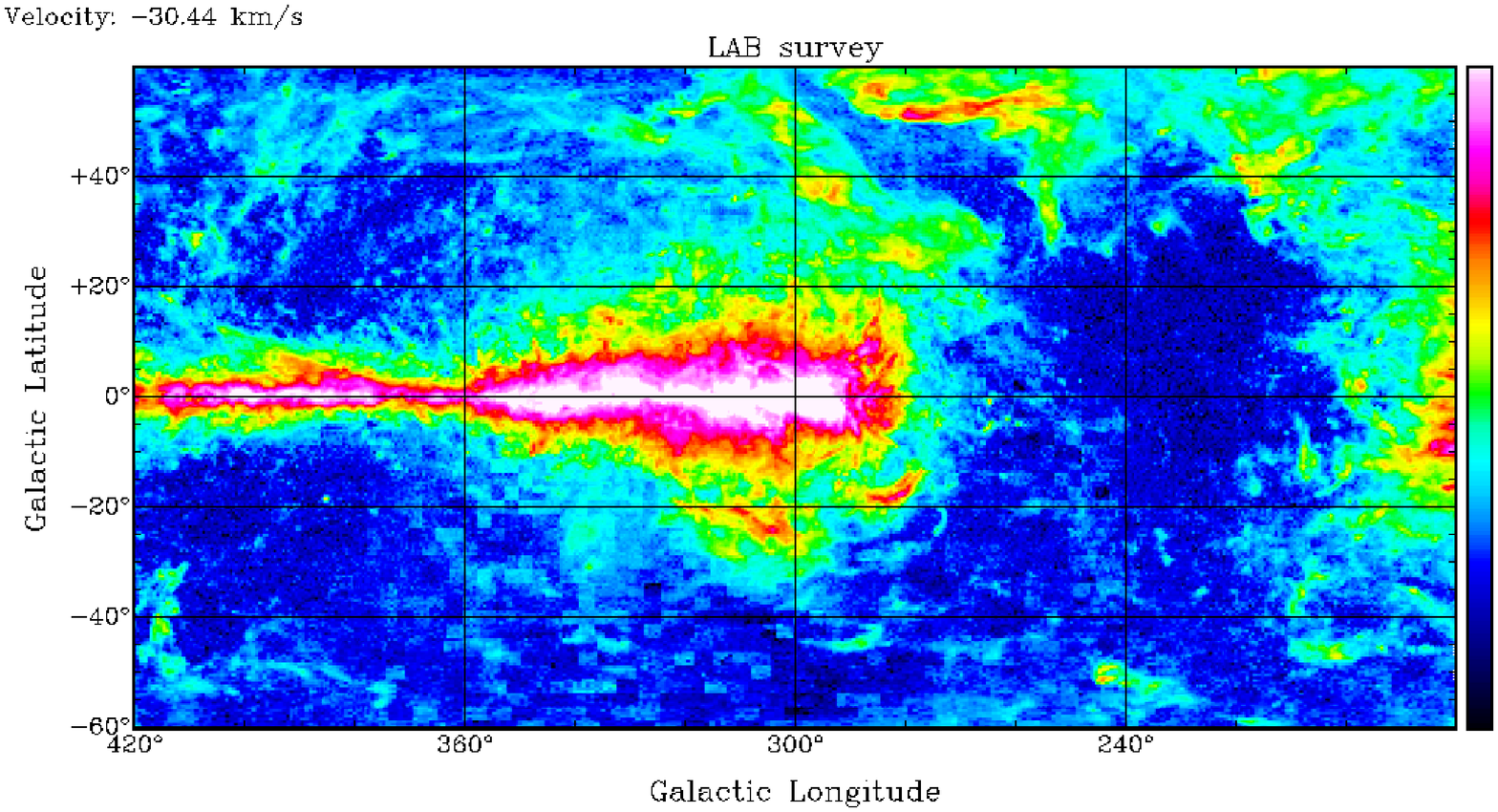}
   \caption{HI emission at ${\rm v_{lsr}} = -30.5 $ \kms. The intensity scale
     for the range $-0.12 < T < 50$ K is logarithmic, emphasizing low level
     emission. Top: GASS, corrected for instrumental baseline but without
     stray-radiation correction, middle: GASS corrected for stray-radiation
     and instrumental baseline, bottom: LAB survey for comparison, showing
     faint patchy spurious features that do not appear in the new GASS data,
     e.g. at $l = 335\deg$, $b = -40\deg$.
   }
   \label{Fig_Map1}
\end{figure*}
%-----------------------------------------------------------------------

%-----------------------------------------------------------------------
\begin{figure*}[!t]
   \centering
   \includegraphics[angle=-90,width=14.5cm]{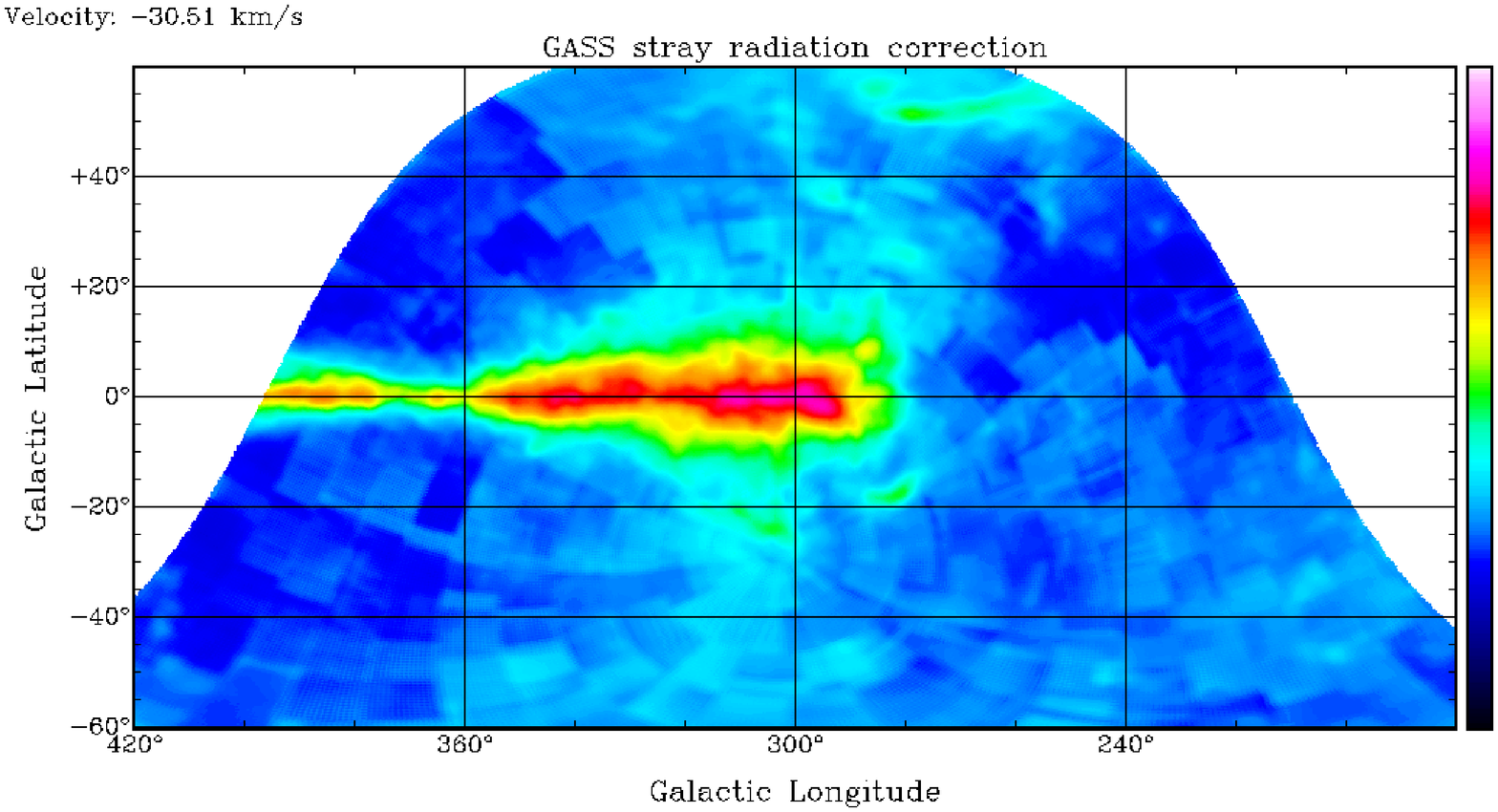}
   \includegraphics[angle=-90,width=14.5cm]{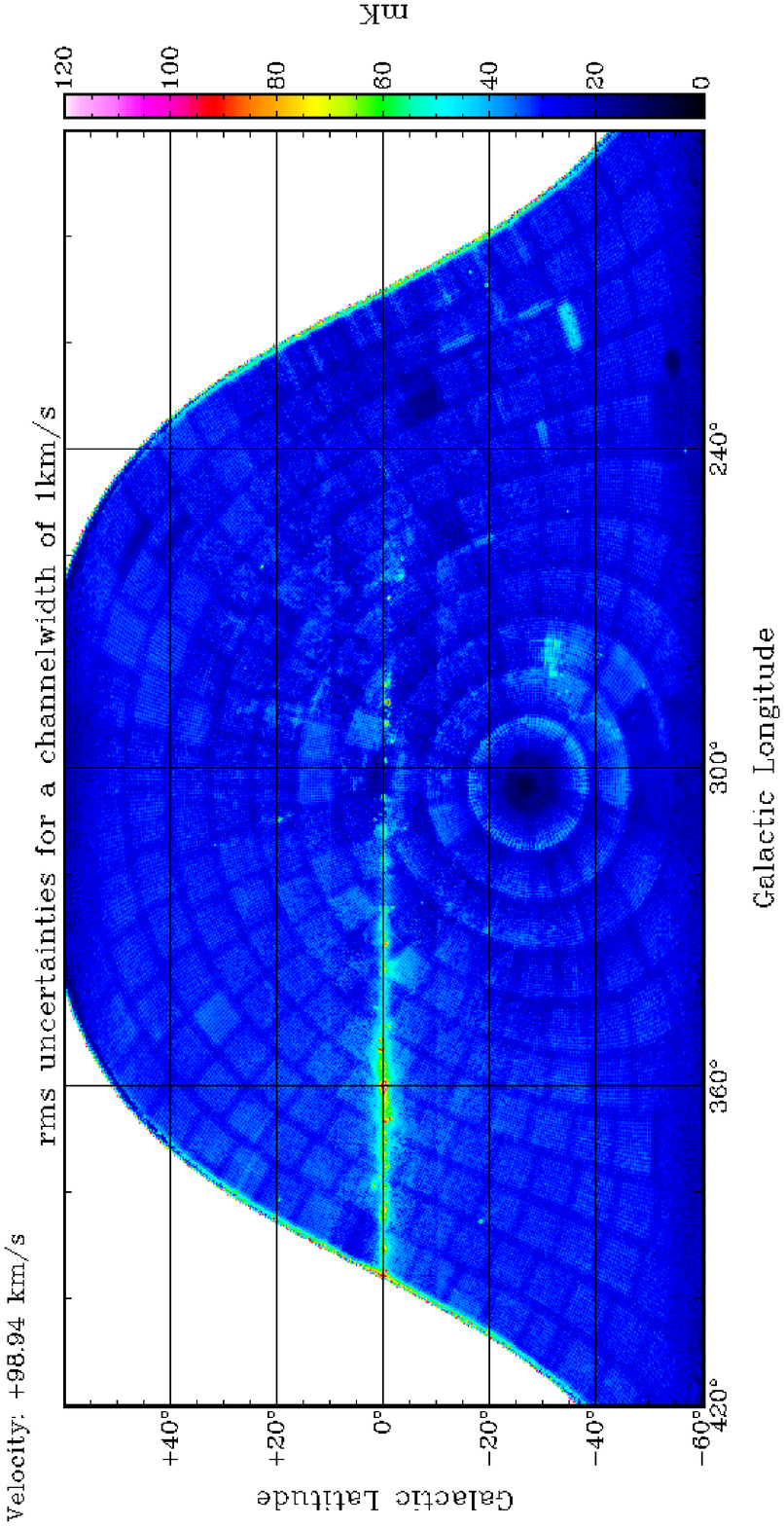}
   \caption{Top: Stray-radiation contribution removed from the GASS at ${\rm
       v_{lsr}} = -30.5 $ \kms, the intensity scale is identical to
     Fig. \ref{Fig_Map1}. Bottom: rms noise in the final GASS data,
     corresponding to the intensity map shown in the central panel of
     Fig. \ref{Fig_Map1}.  }
   \label{Fig_Map2}
\end{figure*}
%-----------------------------------------------------------------------

The noise level of the output map depends on the user defined smoothing
kernel. We measured the noise in the same FITS cubes that we used to fit
beamwidths, at a velocity of 240 \kms~ where neither \hi~ emission nor RFI is
visible. Figure \ref{Gass_noise} gives the results. The noise strongly depends
on the smoothing kernel; for $W_k = 7\farcm0$ we obtain an rms brightness
temperature noise of 57 mK, comparable to the results in
\citetalias{Naomi2009}.  The corresponding effective resolution is $W =
15\farcm6$. 

The noise level depends on the strength of the line emission. Within the
Galactic plane for $|b| \la 1\deg$ we find a typical increase by a factor of
1.75. This approximate upper limit is plotted in Fig. \ref{Gass_noise}. For a
detailed map of the uncertainties we refer to the bottom map in
Fig. \ref{Fig_Map2}. In this case $W_k = 12\farcm0$ was chosen, resulting in a
typical noise level of 35 mK. It is obvious that the noise in the final maps
is affected by RFI mitigation and observational setup. Overlap regions for the
raster chosen are clearly visible.

We tested also the optimal gridding of single dish on-the-fly observations as
proposed by \citet{Mangum2007} with the function
\begin{equation}
  \frac{J_1(r/a)}{(r/a)} \exp{-(r/b)^2}. 
\label{EQBessel}
\end{equation}
$J_1$ is the Bessel function, $a = 1.55$, and $b = 2.52$ for a grid of 1/3
FWHM. This is the default for the AIPS task SDGRD \citep{Greisen1998}. The
resulting resolution and noise are plotted as horizontal lines in
Figs. \ref{Gass_res} and \ref{Gass_noise}. Both median gridding according to
\citetalias{Naomi2009} and optimal Bessel filtering are consistent with a
Gaussian gridding for a kernel of $W_k \sim 7\farcm5$.

\section{Quality of the Data}
\label{Results}

To aid in the discussion of this release of the GASS data we have chosen a
channel map at ${\rm v_{lsr}} = -30.5 $ \kms\ (Fig. \ref{Fig_Map1}). At this
velocity stray-radiation effects are strong and residual problems easily
visible.  In the top panel of Fig. \ref{Fig_Map1}, we show a map that was
generated without a stray-radiation correction but with the instrumental
baselines removed as described in Sect. \ref{Baseline}. In the middle panel,
we plot the map from the final GASS database, corrected for stray-radiation
and instrumental baseline. We have chosen a FWHM kernel $W_k = 12\farcm$,
resulting in rms uncertainties of $\sim 35$ mK at intermediate Galactic
latitudes. In the bottom panel of Fig. \ref{Fig_Map1}, we show the same
channel from the LAB database for comparison. Figure \ref{Fig_Map2} (top)
contains the stray-radiation contribution removed from the middle map in
Fig. \ref{Fig_Map1}.

The bottom panel of Fig. \ref{Fig_Map2} displays a compilation of the residual
$1 \sigma$ rms uncertainties for a FITS cube generated with FWHM kernel $W_k =
12\farcm$ These uncertainties depend on the integration time and on the system
noise. Enhanced noise is caused by low elevation observations, by continuum
emission from the Galactic plane, or by RFI mitigation (rejecting data).

The final GASS map, Fig. \ref{Fig_Map1} middle, shows no obvious evidence of
residual stray-radiation problems. Time dependence of stray-radiation would
cause a correlation with the observational grid that is easily visible in
Fig. \ref{Fig_Map2} (bottom). Such residual problems can be seen in the LAB
channel map (Fig. \ref{Fig_Map1} bottom). Individual blocks, corresponding to
a grid of $ 5 \times 5$ positions, show up at latitudes of $b \sim -50\deg$, a
problem already mentioned by \citet{Kalberla2005}. Similar block structures
existed also in the first version of the Leiden/Dwingeloo Survey
\citep{Atlas1997,Hartmann1996} but could be removed by an improved
stray-radiation correction in the second data release of the LAB
\citep{Kalberla2005}. Time variability is caused predominantly by a seasonal
variation of the stray-radiation and needs to be minimized by the correction
\citep{Kalberla1980,Kalberla1980s}.  The final GASS database is much less
affected by residual instrumental problems than the LAB survey. This implies
in particular that there was no error propagation from the LAB to the GASS. We
conclude that our corrections successfully removed most of the sidelobe
influences, remaining systematical errors are probably below 1\%. The
performance of the survey is equivalent to observations with a telescope
having $\ga $99\% main beam efficiency. Scale uncertainties, caused by gain
fluctuations, are also probably below 1\% (Sect. \ref{Cal}).

The top panel of Fig.~\ref{Fig_Map1}, shows that stray-radiation can
not be removed with only our baseline correction algorithm as most of the
stray-radiation remains in the map.  This also demonstrates that our
baselining algorithm is safe in the sense that it does not noticeably affect
extended profile wings.

We searched for remaining systematical uncertainties in the instrumental
baseline. Some residual uncertainties caused by RFI are visible in individual
channel maps. In most cases such defects remain on average below 40 mK but
there are a few more serious cases. In critical cases we recommend comparing
GASS data with the LAB database. Also a comparison of the cleaned database
with the dirty data, offered by the web interface, may be helpful.

\section{Data products}
\label{Data}
Data from the first release are accessible at
http://www.atnf.csiro.au/research/GASS/Data.html. We provide a web interface
at http://www.astro.uni-bonn.de/hisurvey for the second release to retrieve
spectra and column densities for individual positions or complete FITS
cubes. The user can specify Galactic or Equatorial coordinates, the velocity
range, and the FWHM of the smoothing kernel for the optimal resolution.
Default values for the gridding are provided as well. Some restrictions to the
sizes of the FITS cubes apply, and we currently provide maps only in Cartesian
projection.

The user can check the data for residual contaminations by RFI or
stray-radiation by downloading FITS cubes with dirty data (corrected for
baseline and stray-radiation but no RFI filtering applied) or stray-radiation
corrections that were subtracted as described in Sect. \ref{SR}. Data at high
velocities $|v_{\rm lsr}| \ga 160 $ \kms~ can be generated alternatively from
both IFs or from single ``ghost-free'' IFs only (see Sect. \ref{ghosts}). In
the latter case the rms noise for channels at velocities $|v_{\rm lsr}| \ga
160 $ \kms~ increases by a factor of $\sqrt{2}$.

Despite all efforts some instrumental problems may still exist in the
second data release. Our web based policy for the on-the-fly
generation of FITS cubes allows appropriate corrections. We ask users
of the GASS data to report problems.  Information on updates of the
database will be provided on the web page.

\section{Summary}

The Parkes Galactic All-Sky Survey (GASS) measured the \hi~ emission in the
Southern sky with complete sampling at all declinations $\delta \leq1\deg$
using the Parkes Radioe Telescope. The survey has an effective angular
resolution of $14\farcm4$ at a velocity resolution of $1.0$ \kms, and the data
were obtained in a way that is sensitive to extended diffuse emission as well
as compact sources.  The GASS database contains $2.8 \cdot 10^7$ individual
spectra with 5 second integration time observed by on-the-fly mapping.  The
first data release together with a detailed description of the survey goals
and techniques was given in \citetalias{Naomi2009}. Here we focus on the post
processing to correct instrumental effects including stray-radiation, radio
frequency interference, and residual instrumental spectral baselines.

The GASS is the most sensitive, highest angular resolution survey of Galactic
\hi~ emission ever made of the Southern sky. After correcting for instrumental
effects and RFI the GASS data are in excellent agreement with the LAB
survey. This is impressively demonstrated with Fig. \ref{Fig_Map1}.  FITS data
cubes are available on request from http://www.astro.uni-bonn.de/hisurvey.
Our software allows a flexible generation of FITS datacubes with effective
resolutions $ 16 \la {\rm FWHM} \la 25\arcmin$ (Fig. \ref{Gass_res}). The
corresponding rms noise fluctuations at high Galactic latitudes are in the
range $60 \ga \sigma \ga 20$ mK (Fig. \ref{Gass_noise}).  We provide clean
maps as the final product from our data processing but also maps without RFI
elimination. To allow estimates on possible residual stray-radiation effects
it is also possible to extract the corrections which were applied.

% ======================================================================
% ======================================================================

\begin{acknowledgements} 
  We thank Butler Burton for a very thorough refereeing of this paper with
  constructive criticism and many valuable suggestions. The GASS would have
  not been possible without support from the ATNF staff at Parkes and
  Sydney. We greatly acknowledge assistance in determining the properties of
  the antenna pattern from M. Kesteven, M. Price, J. Reynolds, and
  W. Wilson. P. M\"uller (MPIfR Bonn) provided software for baseline fitting.
  This project was supported by Deutsche Forschungsgemeinschaft, DFG grant
  KA1265/5-2. PK acknowledges also support during a productive stay in Sydney
  as a distinguished visitor of the ATNF. The Parkes Radio Telescope is part
  of the Australia Telescope which is funded by the Commonwealth of Australia
  for operation as a National Facility managed by CSIRO. DJP acknowledges
  partial support for this project from NSF grant AST0104439 and from an
  international faculty development grant from West Virginia University.  He
  thanks the ATNF for its generosity and hospitality during his visit in
  November 2009 to work on this paper.  The National Radio Astronomy
  Observatory is operated by Associated Universities, Inc., under a
  cooperative agreement with the National Science Foundation.
\end{acknowledgements}

\end{document}